\documentclass[reprint, amsmath,amssymb, aps]{revtex4-2}

\usepackage{graphicx}
\usepackage{dcolumn}
\usepackage{bm}
\usepackage{epsfig}
\usepackage{color}
\usepackage{bbm}
\usepackage{longtable}
\usepackage{graphicx}
\usepackage{dcolumn}
\usepackage{bm}

\def\ketm#1{  \left\vert  #1   \right\rangle   }

\def\bram#1{  \left\langle  #1   \right\vert   }

\def\sprm#1#2{  \left\langle #1 \left\vert \right. #2 \right\rangle   }

\def\mem#1#2#3{  \left\langle #1 \left\vert  #2 \right\vert #3 \right\rangle   }

\begin{document}

\preprint{}
%
%
%
%
\title{Radiative recombination of twisted electrons with hydrogen--like heavy ions:\\[0.2cm] Linear polarization of emitted photons}
%
%
%
%

\author{Anna V. Maiorova}
\affiliation{Center for Advanced Studies, Peter the Great St. Petersburg Polytechnic University, 195251 St. Petersburg, Russia}

\author{Anton A. Peshkov}
\affiliation{Physikalisch--Technische Bundesanstalt, D--38116 Braunschweig, Germany}
\affiliation{Institut f\"ur Mathematische Physik, Technische Universit\"at Braunschweig, D--38106 Braunschweig, Germany}

\author{Andrey Surzhykov}
\affiliation{Physikalisch--Technische Bundesanstalt, D--38116 Braunschweig, Germany}
\affiliation{Institut f\"ur Mathematische Physik, Technische Universit\"at Braunschweig, D--38106 Braunschweig, Germany}
\affiliation{Laboratory for Emerging Nanometrology Braunschweig, D-38106 Braunschweig, Germany}

\date{\today \\[0.3cm]}

%
%
%
%

\begin{abstract}
We present a theoretical investigation of the radiative recombination of twisted Bessel electrons with initially hydrogen--like (finally helium--like) heavy ions. In our study, we focus especially on the linear polarization of x--ray photons emitted in the electron capture into the ground $1s^2_{1/2}$ ionic state. Particular emphasis is placed on the question of how this polarization is affected if incident hydrogen--like ions are themselves spin--polarized. To explore such a ``polarization transfer'' we apply the density matrix theory and derive the Stokes parameters of recombination x--rays for the realistic case of collisions between macroscopic electron and ion beams. Based on the developed general approach two scenarios are discussed that are of interest for the planned experiments at ion storage rings. First, we demonstrate how the use of twisted electrons can empower the known method for the diagnostics of spin--polarized ion beams, based on the rotation of the linear polarization of recombination light. In the second scenario we show how the internal structure of ions beams with inhomogeneous intensity and spin patterns can be probed by the capture of Bessel electrons, carrying different values of angular momentum. 
\end{abstract}

\maketitle

%
%
\section{Introduction}

The radiative recombination (RR) of a free electron into a bound state of highly--charged heavy ion has been in the focus of intense research for many years \cite{PaS92,EiM95,FrS05,EiS07,SuJ06}. Such interest stems from the fact that the RR is one of the dominant processes in relativistic ion--electron as well as ion--atom collisions at storage rings. Moreover, it can be seen as the time--reversed photoionization and hence provides an access to the fundamental process of light--matter interaction under the critical conditions of relativistic energies and extremely strong fields. To gain more insight into the radiative recombination a large number of experiments have been performed during last decades at the GSI storage ring in Darmstadt. While in the first experiments the total RR cross sections were mainly measured, more recent studies have dealt with the angular distributions and even \textit{linear polarization} of emitted recombination photons \cite{StL99,StM01,TaS06,VoW19}. The latter, polarization--resolved, measurements became feasible due to the progress in the development of segmented Ge and Si(Li) detectors \cite{SpB08,VoW17}. Such detectors exploit the polarization sensitivity of the Compton effect and allow high--precision determination of both, the degree and direction of linear polarization of RR photons with energies ranging from tens to hundreds of keV's. 

The RR polarization measurements, combined with the rigorous calculations based on Dirac's theory, can reveal unique knowledge about relativistic, many--body, magnetic interaction or even quantum electrodynamics (QED) effects in the electron--photon coupling in high--energy domain \cite{EiS07,TaS06,SuA11,HoA15}. Moreover, the linear polarization of recombination photons provides a very effective tool for the diagnostics of spin states of ion beams in storage rings. It was found, in particular, that the longitudinal polarization of (initially) hydrogen--like heavy ions results in the \textit{rotation} of the linear polarization of the $K$--shell recombination photons out of the reaction plane \cite{SuF05,SuA11}. This ``polarization--rotation'' approach for the beam diagnostics has attracted particular interest since the operational access and control of polarized heavy ions are highly demanded for a variety of atomic collision experiments which are planned both at the GSI and FAIR facilities in Dramstadt and at the Gamma Factory in CERN \cite{BuC20}. For instance, beams of polarized hydrogen--like ions are essential for investigating spin--flip atomic transitions \cite{KlA02} and parity--violation phenomena \cite{BoP11} in high--$Z$ domain as well as for the search of a permanent electric dipole moment.

Until now, the use of the radiative recombination as a probe process for the diagnostics of spin--polarized ions beams has been discussed exclusively for conventional \textit{plane--wave} electrons. Recent years have witnessed, however, remarkable progress in the production and application of \textit{twisted}, or vortex, electron states \cite{BlB07,UcT10,LlB17,BlI17}. Such electrons can carry a non--zero projection of the orbital angular momentum (OAM) onto their propagation direction and possess a helical wave front. The angular momentum, as an additional degree of freedom, makes twisted electrons a powerful tool for probing magnetic phenomena in solid--state physics \cite{GrH17,BlI17} and light--matter coupling \cite{IvK13,SiZ18} as well as for studying various atomic scattering processes \cite{KaK15,MaF18,GoZ20}. We can expect that OAM--effects will also play an important role in the fundamental process of the radiative recombination and can significantly modify the linear polarization of emitted photons. A first step towards such an analysis of the RR linear polarization has been done by Zaytsev and co--workers \cite{ZaS17} who considered the capture of twisted electrons by bare ions. As mentioned above, however, much of today's interest is focused on the radiative recombination of initially hydrogen--like (finally helium--like) ions, whose spin states can be studied based on the polarization--resolved measurements of $K$--RR radiation. 

In the present work, we report a theoretical investigation of the radiative recombination of twisted Bessel electrons with initially hydrogen--like heavy ions. Special attention in our study is paid to the linear polarization of $K$--RR photons and to its sensitivity to the spin state of ions. The analysis of this ``polarization sensitivity'' requires the knowledge of the matrix elements that describe the radiative transition of an electron between continuum and bound ionic states. In Sec.~\ref{subsec:matrix_element} the independent particle model is employed to express these (two--electron) matrix elements in terms of their single--particle counterparts, whose evaluation is well known from the previous studies. With the help of the RR matrix elements we derive later the density matrix of $K$--RR photons. In Sec.~\ref{subsec:density_matrix} we show that this density matrix depends also on the polarization of initially hydrogen--like ions which is described by a set of the so--called statistical tensors. It is exactly the dependence that opens up a way for the \textit{spin--diagnostics} of ion beams, based on the analysis of the linear polarization of RR photons. To perform such a polarization analysis it is practical to construct the Stokes parameters of light from the elements of the photon density matrix. These Stokes parameters for the $K$--RR photons are derived in Sec.~\ref{subsec:linear_polarization} for the realistic experimental scenario of collisions between high--intensity electron and heavy ion \textit{beams}. Since in the present storage ring experiments the radii of both beams is about few millimeters, the polarization parameters have to be obtained upon the averaging over the electron--ion impact parameters ${\bm b}$, as discussed in Sec.~\ref{subsec:b_averaged_density_matrix}. While the derived formulas are general and can be applied to analyze the linear polarization of photons emitted in the $K$--RR of Bessel electrons with the beams of \textit{any} intensity and spin structure, two particular cases are discussed in the present work. In Sec.~\ref{subsec:nonstructured_beams}, for example, we consider the beam of longitudinally polarized  hydrogen--like ions in their ground state. We assume that this polarization, i.e. preferred population of a particular magnetic substate $\ketm{1s_{1/2} \, \mu_0}$, is independent of the position of an ion in the beam, which is usual for the present storage--ring experiments. Our calculations indicate that the polarization of such \textit{unstructured} beams leads to the rotation of the linear polarization of $K$--RR photons out of the reaction plane. Moreover, the rotation angle can be significantly enhanced in a controllable way and, hence, easier measured by modern polarization detectors, if Bessel electrons in place of plane--wave ones are used in experiment. The second scenario is discussed later in Sec.~\ref{subsec:structured_beams} for \textit{structured} ion beams, whose (local) intensity and polarization are varying within their cross--sectional area. Again, we show that detailed information about the internal structure of such beams can be obtained by studying the $K$--RR of Bessel electrons with various projections of angular momentum and by measuring the tilt angle of emitted radiation. The summary of these results and the outlook are presented finally in Sec.~\ref{sec:summary}. 

Relativistic units ($\hbar = c = m_e = 1$) are used throughout the paper unless stated otherwise.

%
%
\section{Theoretical background}
\label{sec:theory}

\subsection{Evaluation of the transition amplitude}
\label{subsec:matrix_element}

The theoretical analysis of the radiative recombination of electrons with highly--charged ions is usually based on the perturbative treatment of the coupling to the electromagnetic radiation \cite{EiM95,EiS07}. In this approach all the properties of the emitted photons can be expressed in terms of the first--order RR amplitude. For the capture of a twisted electron into the ground state $1s^2_{1/2}$ of a finally helium--like ion this amplitude reads as:
\begin{eqnarray}
    \label{eq:amplitude_He}
    {\mathcal T}^{\rm (tw)}_{m_j m_s \lambda \mu_0}({\bm b}) && \nonumber \\
    && \hspace*{-0.5cm}
    = \mem{1s^2_{1/2}: \, {}^1S_0}{\hat{R}_\lambda^\dag}{1s_{1/2} \, \mu_0, \varkappa m_j p_z m_s} \, .
\end{eqnarray}
Here, $\hat{R}_\lambda$ is the transition operator that describes the interaction of the electrons with the radiation field. Within the Coulomb gauge and the relativistic framework, it can be written 
\begin{equation}
    \label{eq:R_operator}
    \hat{R}_\lambda =  \frac{-e}{\sqrt{2\omega(2\pi)^3}}  \, \sum\limits_{q=1,2} {\bm \alpha}_q \, {\bm \epsilon}_{{\bm k} \lambda} \, e^{i{\bm k} {\bm r}_q}
\end{equation}
as a sum of one--particle operators, where ${\bm r}_q$ is the position vector and ${\bm \alpha}_q$ is the vector of Dirac matrices for the $q$--th electron. Moreover, the emitted photon is characterized by its energy ${\omega}$,  wave-- and (circular) polarization vectors, ${\bm k}$ and ${\bm \epsilon}_{{\bm k} \lambda}$, with $\lambda = \pm 1$ being the photon's helicity.   

In order to further evaluate the transition amplitude (\ref{eq:amplitude_He}), we have to agree how to describe the system ``ion + electrons'' before and after the radiative recombination takes place. The initial state of the system is given by an ion in the ground hydrogenic state $\ketm{1s_{1/2} \, \mu_0}$, with $\mu_0 = \pm 1/2$ being the projection of the total angular momentum, \textit{and} by a continuum electron incident along the $z$--axis. Below we will assume that this electron is prepared in the twisted Bessel state $\ketm{\varkappa m_j p_z m_s}$, which is characterized by the well--defined projections $p_z$ and $m_j$ of the linear-- and total angular momenta, as well as by the fixed absolute value of the transverse momentum $\varkappa = \left| {\bm p}_\perp \right|$. Since the properties of the Bessel electrons, moving in the Coulomb field of an ion, have been thoroughly discussed in the literature \cite{ZaS17,GoZ20,MaP20}, here we just briefly recall the basic expressions, required for the further analysis. The wave function of an electron in the Bessel state can be written as the superposition of the plane--wave continuum solutions $\psi_{{\bm p} m_s}({\bm r})$ of the Dirac equation:
\begin{eqnarray}
    \label{eq:wave_function_Bessel}
    \psi_{\varkappa m_j p_z m_s}({\bm r}) &=& \int{{\rm d}{\bm p} \, a_{\varkappa m_j}({\bm p}) \, 
    {\rm e}^{i {\bm p} {\bm b}} \, \psi_{{\bm p} m_s}({\bm r})} \, .
\end{eqnarray}
Here, the amplitude $a_{\varkappa m_j}({\bm p})$ is given by:
\begin{eqnarray}
    \label{eq:a_amplitude}
    a_{\varkappa m_j}({\bm p}) &=& i^{m_s - m_j} \, \frac{{\rm e}^{i m_j \varphi_p}}{2 \pi p_\perp} \, \delta\left(p_{\parallel} - p_z\right) \, \delta\left(p_{\perp} - \varkappa\right) \, ,
\end{eqnarray}
and the position of an electron with respect to a target atom is specified by the impact parameter ${\bm b} = \left(b_x, b_y, 0 \right)$. The need to introduce this parameter stems from the inhomogeneous internal structure of Bessel electron states. In particular, both the probability density distribution of the electrons and their spin polarization varies with the distance from the center of the Bessel beam, see Refs.~\cite{ZaS17,GoZ20,MaP20} for further details. 

While before the RR one electron is in the bound state and the other is in the continuum state, both of them form the ground state $1s^2_{1/2}$ of the helium--like ion \textit{after} the recombination. In general, accurate theoretical description of such a two--electron state is a rather complicated task which requires an account of the $e$--$e$ interaction. For the high--$Z$ domain, however, the interelectronic interaction effects are rather small comparing to the electron--nucleus coupling, and the state $\ketm{1s^2_{1/2}: \, {}^1S_0}$ can be approximated within the framework of the independent particle model (IPM). This model, which takes the Pauli exclusion principle into account, has been successfully applied in the past to describe the radiative recombination into finally helium--like ions \cite{SuJ08,SuA11}. A great advantage of the IPM approach is that it allows to express two--electron recombination amplitudes in terms of their one--electron counterparts. For example, Eq.~(\ref{eq:amplitude_He}) can be expressed as:
\begin{eqnarray}
    \label{eq:amplitude_He_2}
    {\mathcal T}^{\rm (tw)}_{m_j m_s \lambda \mu_0}({\bm b}) 
    = (-1)^{1/2 + \mu_0} \, \tau^{\rm (tw)}_{m_j m_s \lambda - \mu_0}(\bm b) \, ,
\end{eqnarray}
where the amplitude $\tau^{\rm (tw)}_{m_j m_s \lambda - \mu_0}(\bm b)$ describes the capture of an electron into a \textit{single--particle} substate $\ketm{1s_{1/2} -\mu_0}$. The minus sign in front of $\mu_0$ simply follows from the Pauli principle and from the fact that the other substate, 
$\ketm{1s_{1/2} \, \mu_0}$, is already occupied by a spectator electron. Not so much has to be said about further evaluation of $\tau^{\rm (tw)}_{m_j m_s \lambda - \mu_0}(\bm b)$ for the case of incident Bessel electrons. As it was already shown in Refs.~\cite{ZaS17,MaP20}, this ``twisted'' amplitude can be written as:
\begin{eqnarray}
    \label{eq:tau_amplitude_twisted}
    \tau^{\rm (tw)}_{m_j m_s \lambda -\mu_0}(\bm b) && \nonumber \\
    && \hspace*{-1cm} = {\rm e}^{-i {\bm k} {\bm b}} \,
    \int{{\rm d}{\bm p} \, a_{\varkappa m_j}({\bm p}) \, {\rm e}^{i {\bm p} {\bm b}} \,
    \tau^{\rm (pl)}_{m_s \lambda -\mu_0}({\bm p}) } \, ,
\end{eqnarray}
where $a_{\varkappa m_j}({\bm p})$ is given by Eq.~(\ref{eq:a_amplitude}), and $\tau^{\rm (pl)}_{m_s \lambda -\mu_0}({\bm p})$ is the amplitude for the radiative recombination of the usual plane--wave electron with the asymptotic momentum ${\bm p}$ and spin projections $m_s$.

\subsection{Density matrix approach}
\label{subsec:density_matrix}

Having briefly discussed the evaluation of the amplitude (\ref{eq:amplitude_He}) for the RR of the twisted electron into the ground state of helium--like ions, we are ready to investigate the polarization of emitted photons. Most conveniently this can be done within the framework of the density matrix theory. Again, the application of this theory to the capture of (plane--wave) electrons into bound states of highly--charged ions has a long successful history and was discussed elsewhere \cite{SuF02,SuJ06,EiS07}. In the present work, therefore, we will omit the details of the derivation and proceed directly to the density matrix of emitted photons:
\begin{widetext}
\begin{eqnarray}
    \label{eq:photon_density_matrix_1}
    \mem{{\bm k} \lambda}{\hat{\rho}_{\gamma}({\bm b})}{{\bm k} \lambda'} &=& {\mathcal N} \; \sum\limits_{\mu_0 \mu'_0 m_s}
    {\mathcal T}^{\rm (tw)}_{m_j m_s \lambda \mu_0}({\bm b}) \, {\mathcal T}^{\rm (tw) *}_{m_j m_s \lambda' \mu'_0}({\bm b}) \, \mem{1s_{1/2} \, \mu_0}{\hat{\rho}_{0}({\bm b})}{1s_{1/2} \, \mu'_0}  \nonumber \\[0.2cm]
    &=& - {\mathcal N} \; \sum\limits_{\mu_0 \mu'_0 m_s} (-1)^{\mu_0+\mu'_0} \; \tau^{\rm (tw)}_{m_j m_s \lambda - \mu_0}({\bm b}) \, \tau^{\rm (tw) *}_{m_j m_s \lambda' - \mu'_0}({\bm b}) \,  \mem{1s_{1/2} \, \mu_0}{\hat{\rho}_{0}({\bm b})}{1s_{1/2} \, \mu'_0} \, ,
\end{eqnarray}
\end{widetext}
where in the second line we have employed the independent particle model (\ref{eq:amplitude_He_2}) to evaluate the RR amplitude. The factor ${\mathcal N}$ is introduced in order to the density matrix has a usual normalization (${\mathrm{Tr}\left(\hat{\rho}_{\gamma}\right) = 1}$). We have assumed, moreover, that while the incident Bessel electrons carry a well--defined value of the total angular momentum $m_j$, their spin projection $m_s$ is not fixed.

As seen from Eq.~(\ref{eq:photon_density_matrix_1}), the elements of the photon density matrix, written in the helicity representation $\ketm{{\bm k} \lambda}$, can be expressed in terms of the transition amplitudes $\tau^{\rm (tw)}_{m_j m_s \lambda - \mu_0}({\bm b})$ \textit{and} the initial--state density matrix $\mem{1s_{1/2} \, \mu_0}{\hat{\rho}_{0}({\bm b})}{1s_{1/2} \, \mu'_0}$. The latter describes magnetic sublevel population of the hydrogen--like ions before the electron capture. In our study we will assume that this population may also depend on the position ${\bm b}$ of an ion with respect to the center of the incident Bessel beam. Such a b--dependence can arise, for example, if hydrogen--like ions are themselves produced by the recombination of twisted electrons with bare nuclei \cite{MaP20}.   

To proceed further with the analysis of the polarization properties of recombination photons, it is convenient to re--write the initial--state density matrix in Eq.~(\ref{eq:photon_density_matrix_1}) in terms of the statistical tensors of hydrogen--like ions:
\begin{eqnarray}
    \label{eq:statistical_tensors_1}
    \mem{1s_{1/2} \, \mu_0}{\hat{\rho}_{0}({\bm b})}{1s_{1/2} \, \mu'_0} && \nonumber \\[0.2cm]
    && \hspace{-3.6cm} = 
    \sum\limits_{kq} (-1)^{1/2 - \mu'_0} \, \sprm{1/2 \mu_0 \, 1/2 - \mu'_0}{kq} \, \rho_{kq}({\bm b}) \, .
\end{eqnarray}
The great advantage of $\rho_{kq}$ is that they are constructed to represent the irreducible tensors of rank $k$ and projection $q$ and, hence, obey well--known transformation properties under the rotation of coordinates \cite{FaR59,Blu12}. Moreover, a clear physical interpretation can be given to the components of these tensors. While $\rho_{00}$ is proportional to the probability to find a hydrogen--like ion in a particular state $\ketm{n_0 j_0}$, which is the $1s_{1/2}$ in our case, the orientation and/or alignment of this state  is described by the set of components $\rho_{kq}$ with $k \ne 0$.

By making use of Eq.~(\ref{eq:statistical_tensors_1}) we re--write the density matrix of emitted photons as:
\begin{eqnarray}
    \label{eq:photon_density_matrix_2}
    \mem{{\bm k} \lambda}{\hat{\rho}_{\gamma}({\bm b})}{{\bm k} \lambda'} && \nonumber \\[0.2cm]
    && \hspace*{-2.2cm} = - {\mathcal N} \;\sum\limits_{\mu_0 \mu'_0 m_s} \sum\limits_{kq} \Big[ (-1)^{1/2 - \mu_0} \, \sprm{1/2 - \mu_0 \, 1/2 \mu'_0}{kq} \nonumber\\
    && \hspace*{-1.3cm} \times \, \tau^{\rm (tw)}_{m_j m_s \lambda \mu_0}({\bm b}) \, \tau^{\rm (tw) *}_{m_j m_s \lambda' \mu'_0}({\bm b}) \, \rho_{kq}({\bm b}) \Big] \, . 
\end{eqnarray}
As seen from this expression, the elements of the density matrix $\mem{{\bm k} \lambda}{\hat{\rho}_{\gamma}({\bm b})}{{\bm k} \lambda'}$ depend on the position ${\bm b}$ of an ion within the incident electron wavefront. This $b$--dependence arises from both, the recombination amplitudes and the statistical tensors of the initially hydrogen--like ions. Under realistic experimental conditions, however, the RR is observed in electron and ion \textit{beam} collisions, which do not allow control the impact parameter. Hence, the (observable) properties of the recombination photons, calculated from the density matrix (\ref{eq:photon_density_matrix_2}), have to be averaged over ${\bm b}$ in order to ``fit'' the experimental scenario. This averaging procedure will be discussed in the next section. 

\subsection{Linear polarization of RR photons}
\label{subsec:linear_polarization}

\begin{figure}[t]
	\includegraphics[width=0.95\linewidth]{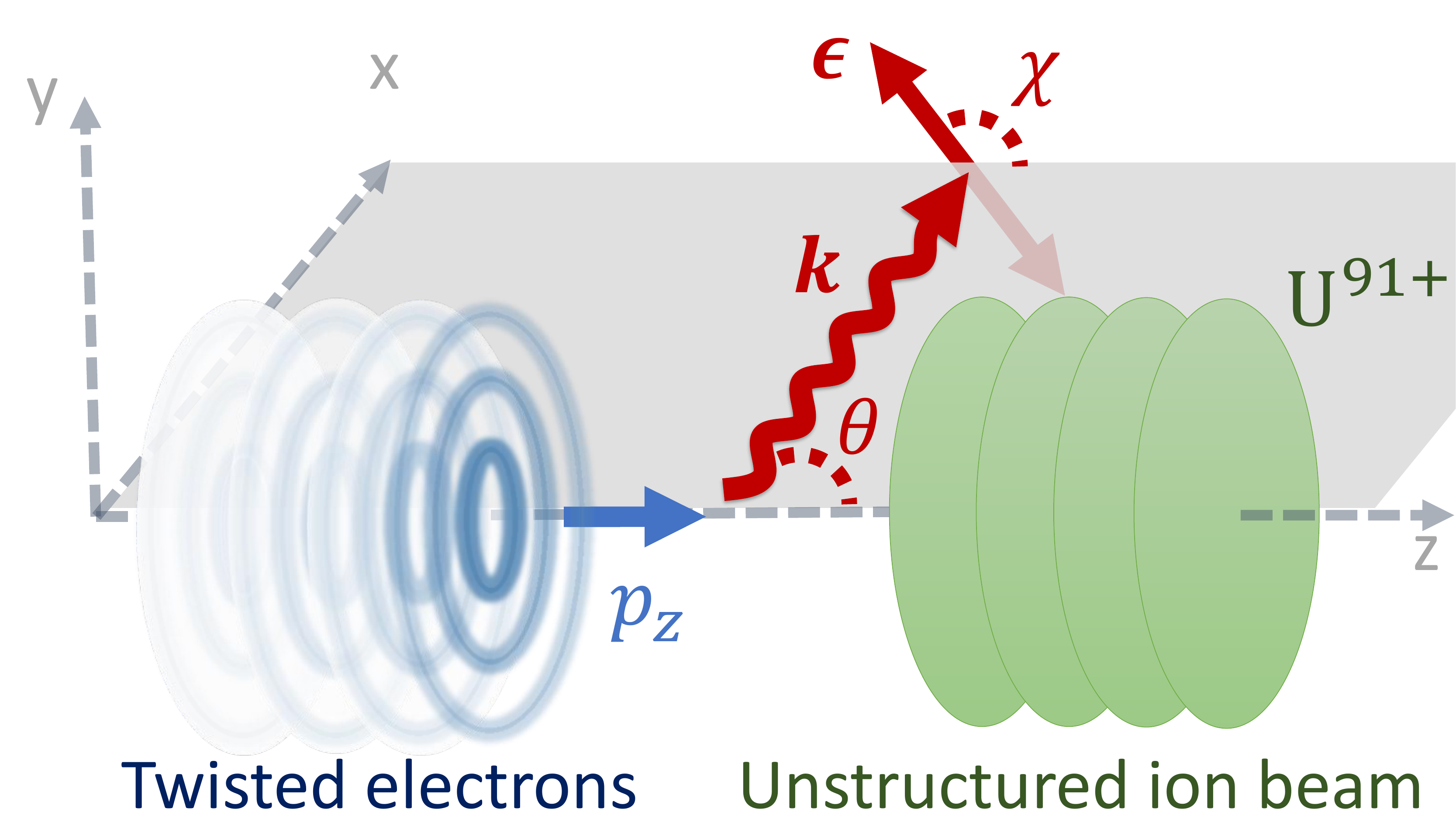}
	\caption{The geometry of the radiative capture of twisted Bessel electrons into the 1s$^2_{1/2}$ ground state of initially hydrogen--like (finally helium--like) uranium ions. The electron beam propagation direction, as ``seen'' in the ion rest frame, is chosen as the $z$ axis. Together with the wave vector ${\bm k}$ of emitted photon this axis defines the scattering ($xz$--) plane. The orientation of the linear polarization of emitted $K$--RR photons is characterized by the tilt angle $\chi$, defined with respect to the scattering plane.}
    \label{Fig1}
\end{figure}

With the help of the density matrix (\ref{eq:photon_density_matrix_2}) one can evaluate all the properties of the photons, emitted in the capture of Bessel electrons into the K--shell of helium--like ions. In the present study we will focus especially on the \textit{linear} polarization of K--RR photons which is usually parameterized by two Stokes parameters $P_1$ and $P_2$. These parameters can be expressed in terms of the intensities of light $I_{\chi}$, linearly polarized under various angles $\chi$ with respect to the \textit{reaction} plane \cite{BaG00,Blu12}. In  experimental and theoretical RR studies this plane is chosen to be spanned by the directions of the incident electron beam ($z$--axis) and of the emitted photons $\hat{\bm k} = {\bm k}/k$, see Fig.~\ref{Fig1}. The intensities of light, polarized within and perpendicular to this plane, define the first Stokes parameter:
\begin{equation}
    \label{eq:Stokes_P1}
    P_1 = \frac{I_{0} - I_{90}}{I_{0} + I_{90}} \, ,
\end{equation}
while $P_2$ follows from a similar expression:
\begin{equation}
    \label{eq:Stokes_P2}
    P_2 = \frac{I_{45} - I_{135}}{I_{45} + I_{135}} \, ,
\end{equation}
obtained for polarization tilt angles $\chi = 45$~deg and $\chi = 135$~deg. 

In order to use Eqs.~(\ref{eq:Stokes_P1})--(\ref{eq:Stokes_P2}) for the computation of the Stokes parameters of RR photons, one has to obtain first the light intensities $I_{\chi}$. Within the framework of the density matrix theory this can be easily achieved by introducing the so--called detector operator:
\begin{eqnarray}
    \label{eq:detector_operator_I}
    \hat{P}_{\chi} &=& \ketm{{\bm k} \, \chi} \bram{{\bm k} \, \chi} \nonumber \\[0.2cm]
    &=& \frac{1}{2} \, \sum\limits_{\lambda \lambda'} {\rm e}^{i \chi (\lambda' - \lambda)} \,
    \ketm{{\bm k} \, \lambda}\bram{{\bm k} \, \lambda'} \, ,
\end{eqnarray}
that describes the ``measurement'' of a photon in the state $\ketm{{\bm k} \, \chi}$ with the wave vector ${\bm k}$ and linear polarization titled by an angle $\chi$ with respect to the reaction plane. As seen from the second line of Eq.~(\ref{eq:detector_operator_I}), this linear polarized state can be expanded in more convenient circular polarization (or helicity) basis, $\ketm{{\bm k} \, \chi} = 1/\sqrt{2} \, \sum_\lambda {\rm e}^{-i \lambda \chi} \ketm{{\bm k} \lambda}$. By taking a trace of the product of the detector operator (\ref{eq:detector_operator_I}) and the photon density matrix (\ref{eq:photon_density_matrix_2}) we find the probability to measure a photon in the state $\ketm{{\bm k} \, \chi}$, which is proportional to the intensity:
\begin{equation}
    \label{eq:intensity_I_1}
    I_{\chi}({\bm b}) = {\mathcal C} \; {\rm Tr}\left(\hat{P}_{\chi} \hat{\rho}_{\gamma}({\bm b}) \right) \, .
\end{equation}
The coefficient ${\mathcal C}$ is independent of the emission and polarization angles of the RR photons and just reflects parameters of a particular experimental setup.   

Eq.~(\ref{eq:intensity_I_1}) describes the intensity of linearly polarized light, emitted from a particular point ${\bm b}$. In nowadays storage--ring experiments, however, the radiative recombination is usually studied in collisions of electron and ion beams, each of which have macroscopic cross--sectional areas. In order to be able to describe the outcome of these experiments we have to average the intensity $I_{\chi}({\bm b})$ over the impact parameter:
\begin{eqnarray}
    \label{eq:intensity_I_2}
    I_{\chi} &=& \int \frac{{\rm d}{\bm b}}{\pi R^2} \, I_{\chi}({\bm b}) \nonumber \\[0.2cm]
    &=& \frac{{\mathcal C}}{2 \pi R^2} \, \sum\limits_{\lambda \lambda'} \,
    {\rm e}^{i \chi (\lambda - \lambda')} \int {\rm d}{\bm b} \mem{{\bm k} \lambda}{\hat{\rho}_{\gamma}({\bm b})}{{\bm k} \lambda'},
\end{eqnarray}
where $R$ is the radius of the ion beam, and Eqs.~(\ref{eq:photon_density_matrix_2}) and (\ref{eq:intensity_I_1}) were employed in the second line of the expression. With the help of this \textit{averaged} intensity $I_{\chi}$ we finally derive the Stokes parameters:
\begin{subequations}
\begin{align}
\label{eq:Stokes_parameters_final_P1}
    P_1 &=& \frac{\sum\limits_{\lambda \lambda'} (1 - \delta_{\lambda,\lambda'}) \, \int {\rm d}{\bm b} \mem{{\bm k} \lambda}{\hat{\rho}_{\gamma}({\bm b})}{{\bm k} \lambda'}}{\sum\limits_{\lambda}\int {\rm d}{\bm b} \mem{{\bm k} \lambda}{\hat{\rho}_{\gamma}({\bm b})}{{\bm k} \lambda}} \, , \\[0.2cm]
    P_2 &=& \frac{i}{2} \, \frac{\sum\limits_{\lambda \lambda'} (\lambda-\lambda') \, \int {\rm d}{\bm b} \mem{{\bm k} \lambda}{\hat{\rho}_{\gamma}({\bm b})}{{\bm k} \lambda'}}{\sum\limits_{\lambda}\int {\rm d}{\bm b} \mem{{\bm k} \lambda}{\hat{\rho}_{\gamma}({\bm b})}{{\bm k} \lambda}} \, ,
    \label{eq:Stokes_parameters_final_P2}
\end{align}
\end{subequations}
that describe the linear polarization of the RR photons, emitted from the \textit{entire} electron and ion beam intersection area.  

\subsection{Evaluation of the $b$--averaged density matrix}
\label{subsec:b_averaged_density_matrix}

As seen from Eqs.~(\ref{eq:Stokes_parameters_final_P1})--(\ref{eq:Stokes_parameters_final_P2}), the Stokes parameters $P_1$ and $P_2$ can be expressed in terms of the averaged elements of the photon density matrix. For the further evaluation of these elements one employs  Eq.~(\ref{eq:photon_density_matrix_2}) as well as the explicit form (\ref{eq:tau_amplitude_twisted}) of the RR amplitude, and finds:
\begin{widetext}
\begin{eqnarray}
    \label{eq:b-averaged_density_matrix}
    \int {\rm d}{\bm b} \mem{{\bm k} \lambda}{\hat{\rho}_{\gamma}({\bm b})}{{\bm k} \lambda'} &=&
    - \mathcal{\widetilde{N}} \;\sum\limits_{\mu_0 \mu'_0 m_s} \sum\limits_{kq} (-1)^{1/2- \mu_0} \, \sprm{1/2 -\mu_0 \, 1/2 \mu'_0}{kq} \nonumber \\
    &\times& \int {\rm d}\varphi_p \, {\rm d}\varphi_{p'} \, {\rm e}^{i m_j (\varphi_p - \varphi_{p'})} \, \tau^{\rm (pl)}_{m_s \lambda \mu_0}({\bm p}) \, \tau^{\rm (pl) *}_{m_s \lambda' \mu'_0}({\bm p}') \, {\mathcal F}_{kq}({\bm p} - {\bm p}') \, .
\end{eqnarray}
\end{widetext}
Here, $\tau^{\rm (pl)}_{m_s \lambda \mu_0}({\bm p})$ is the RR amplitude for the plane--wave electron, whose asymptotic momentum is written in cylindrical coordinates as ${\bm p} = \left({\bm p}_\perp, \, p_\parallel \right) = \left(p_\perp \cos\varphi_p, \, p_\perp \sin\varphi_p, \, p_\parallel \right)$ with longitudinal $p_\parallel$ and tranverse ${\bm p}_\perp$ components defined along the $z$--axis and within the $xy$--plane, respectively. The integration in Eq.~(\ref{eq:b-averaged_density_matrix}) runs only over the azimuthal angle $\varphi_p$, while the absolute values $p_\parallel = p_z$ and $p_\perp = \varkappa$ are fixed to represent the kinematic parameters of the incident Bessel beam, see Eqs.~(\ref{eq:wave_function_Bessel})--(\ref{eq:a_amplitude}). In the calculations below the ratio of these, transverse and longitudinal, components will be parameterized in terms of the so--called opening angle, $\tan\theta_p = \varkappa/p_z$. $\mathcal{\widetilde{N}}$ is the normalization factor  which is not significant for Stokes parameter calculations.

Beside the RR amplitudes, the $b$--averaged density matrix (\ref{eq:b-averaged_density_matrix}) depends also on the Fourier transform of the statistical tensors of the hydrogen--like ions:
\begin{equation}
    \label{eq:Fourier_transform}
    {\mathcal F}_{kq}({\bm p} - {\bm p}') = \int {\rm d}{\bm b} \, {\rm e}^{i({\bm p}- {\bm p}'){\bm b}} \, \rho_{kq}({\bm b}) \, .
\end{equation}
This implies that the polarization Stokes parameters of recombination photons are sensitive to the sublevel population of initial hydrogenic states. In the following section we will investigate this sensitivity for various experimental scenarios.

\section{Results and discussion}
\label{sec:results}

\subsection{Recombination of polarized ion beams}
\label{subsec:nonstructured_beams}

During the last two decades, several schemes have been proposed to produce beams of longitudinally polarized hydrogen--like ions in storage rings. In particular, the optical pumping of hyperfine levels of these ions \cite{PrL03} or the RR of polarized electrons with initially bare ions \cite{BoP11} may lead to a preferred population of a particular hydrogenic substate $\ketm{n_0 \, j_0 \, \mu_0 = \pm j_0}$. For both above scenarios this population is independent of the position ${\bm b}$ of an ion and, hence, one can talk about spin--polarized but spatially \textit{unstructured} beams. Of special interest here are the beams of hydrogen--like ions in their ground 1s$_{1/2}$ state, whose longitudinal polarization is described by the parameter:
\begin{equation}
    \label{eq:H-like_ions_polarization}
    {\mathcal P}_z = \frac{n_{+1/2} - n_{-1/2}}{n_{+1/2} + n_{-1/2}} \, .
\end{equation}
Here, $n_{\pm 1/2} \equiv n_{1s _{1/2} \, \mu_0 = \pm 1/2}$ are the relative populations of magnetic substates $\ketm{1s_{1/2} \, \mu_0}$ and the projection $\mu_0$ of the total angular momentum $j_0 = 1/2$ is defined with respect to the beam direction. 

Beside the production of spin--polarized beams of hydrogen--like ions, the diagnostics of this polarization in the storage ring experiments remains an important task. Recently, the radiative recombination of plane--wave unpolarized electrons into the ground state $1s^2_{1/2}$ of finally helium--like ions has been proposed as a probe process for measuring the degree of ion polarization ${\mathcal P}_z$ \cite{SuF05,SuA11}. This measurement becomes feasible since the linear polarization of $K$--RR photons is sensitive to ${\mathcal P}_z$.  The analysis based on the density matrix theory has indicated, in particular, that the second Stokes parameter of recombination photons is directly proportional to the ion polarization, $P_2 \sim {\mathcal P}_z$, while the $P_1$ is independent of ${\mathcal P}_z$. In order to better ``visualize'' this effect, it is convenient to represent the linear polarization of recombination photons in terms of their polarization ellipse. This ellipse is defined in the plane normal to the photon momentum ${\bm k}$ and its principal axis with the length $P_L = \sqrt{P_1^2 + P_2^2}$ is tilted by the angle:
\begin{equation}
    \label{eq:tilt_angle}
    \tan 2\chi_0 = \frac{P_2}{P_1} \, ,
\end{equation}
with respect to the reaction plane. Based on this (ellipse) representation one concludes that the polarization of initially hydrogen--like ions results in the \textit{rotation} of the linear polarization of $K$--RR photons out of the reaction plane by the angle $\tan 2\chi_0 \sim {\mathcal P}_z$.  

%
%
%
%

Until now, the ``polarization rotation'' method for the ion beam diagnostics has been discussed exclusively for the recombination of unpolarized plane--wave electrons. In the present work we aim to extend it to the case of twisted electrons and to highlight the advantages of this expansion. Similar to the well--elaborated plane--wave case we will consider below the capture of unpolarized Bessel electrons into the ground state of longitudinally polarized hydrogen--like ions. To evaluate the density matrix and the Stokes parameters of $K$--RR photons for this scenario, we have to determine first the statistical tensors $\rho_{kq}$ of initial ions. Since these ions are assumed to be prepared in the ground state 1s$_{1/2}$, their sublevel population is described by just two \textit{non--zero} tensors, $\rho_{00}$ and $\rho_{10}$, which are related to each other by the degree of beam longitudinal polarization:   
\begin{eqnarray}
    \label{eq:stat_tensors_unstructured_beam}
    \rho_{10} = {\mathcal P}_z \, \rho_{00} \, .
\end{eqnarray}
We remind that for the polarized beams, conventionally produced by the optical pumping of hydrogen--like or by the recombination of bare ions with plane--wave electrons, the tensors $\rho_{00}$ and $\rho_{10}$ are independent of the impact parameter ${\bm b}$. For this (unstructured beam) case the Fourier transform (\ref{eq:Fourier_transform}) trivially simplifies to:
\begin{equation}
    \label{eq:Fourier_transform_unstructured_beam}
    {\mathcal F}_{kq}({\bm p} - {\bm p}') = 4 \pi^2 \, \delta\left( {\bm p}_\perp - {\bm p}'_\perp \right) \, \delta_{q0} \,  \, \rho_{k0} \, ,
\end{equation}
with $k =0, 1$, and the $b$--averaged density matrix (\ref{eq:b-averaged_density_matrix}) can be written as:
\begin{eqnarray}
    \label{eq:b-averaged_density_matrix_unstructured}
    \int {\rm d}{\bm b} \mem{{\bm k} \lambda}{\hat{\rho}_{\gamma}({\bm b})}{{\bm k} \lambda'} &&
    \nonumber \\[0.2cm]
    && \hspace*{-3cm} = - \mathcal{\widetilde{N}} \, \sum\limits_{\mu_0 m_s} \sum\limits_{k} (-1)^{1/2 - \mu_0} \, \sprm{1/2 -\mu_0 \, 1/2 \mu_0}{k0} \, \rho_{k0} \nonumber \\
    && \hspace*{-2.5cm} \times \int {\rm d}\varphi_p \, \tau^{\rm (pl)}_{m_s \lambda \mu_0}({\bm p}) \, \tau^{\rm (pl) *}_{m_s \lambda' \mu_0}({\bm p})  \, .
\end{eqnarray}
As seen from this expression, the elements of the density matrix are independent of the projection $m_j$ of the total angular momentum (TAM) of Bessel electrons and just sensitive to the ratio of (the absolute values of) their transverse to longitudinal momenta. This result is consistent with the common knowledge that the probabilities of atomic processes, involving twisted particles, are insensitive to the TAM if the integration over the impact parameter ${\bm b}$ is performed \cite{ZaS17,KaK17}.

%
%
\begin{figure*}[t]
	\includegraphics[width=0.95\linewidth]{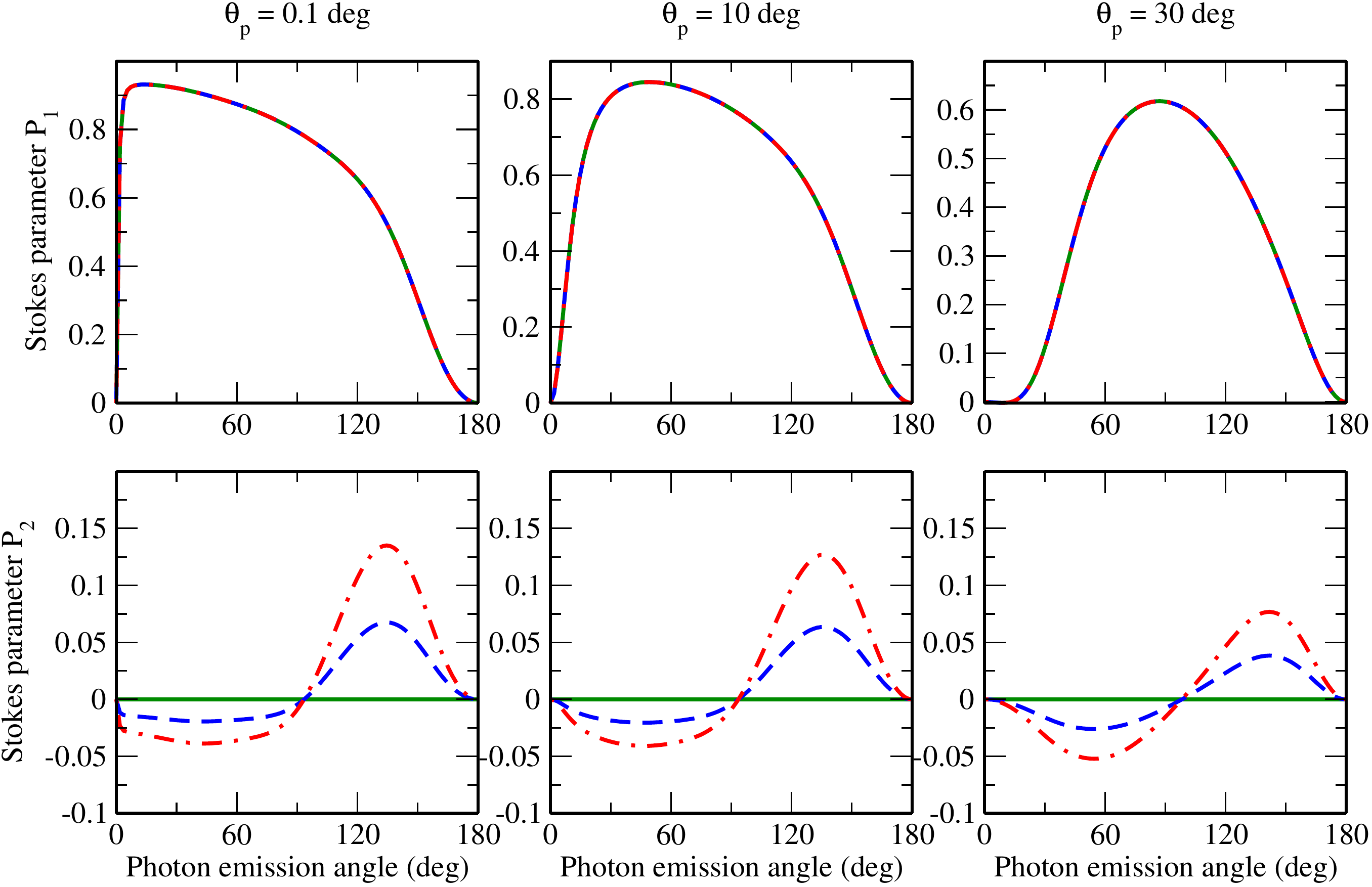}
	\caption{The Stokes parameters $P_1$ (upper panel) and $P_2$ (lower panel) of the $K$--RR photons, emitted in collisions of unpolarized Bessel electrons with an unstructured beam of hydrogen--like uranium U$^{91+}$ ions. Calculations have been performed for the electrons with kinetic energy 100~keV and opening angles $\theta_p$~=~0.1~deg (left column), $\theta_p$~=~10~deg (middle column) and $\theta_p$~=~30~deg (right column). Moreover, the initial--state hydrogen--like ions are assumed to be longitudinally polarized with the degree of polarization ${\mathcal P}_z$~=~0 (green solid line), ${\mathcal P}_z$~=~0.5 (blue dashed line) and ${\mathcal P}_z$~=~1.0 (red dash--dotted line). The results are presented in the ion rest frame.}
    \label{Fig2}
\end{figure*}
%
%

By inserting the density matrix (\ref{eq:b-averaged_density_matrix_unstructured}) into Eqs.~(\ref{eq:Stokes_parameters_final_P1})--(\ref{eq:Stokes_parameters_final_P2}) one can finally obtain the Stokes parameters of the photons, emitted in the $K$--RR of (unpolarized) Bessel electrons with longitudinally polarized hydrogen--like ions. For the sake of brevity we will not present here these---rather complicated---expressions but discuss instead the general properties of $P_1$ and $P_2$ with a special focus on their sensitivity to the ${\mathcal P}_z$. Similar to the recombination of the plane--wave electron we found qualitatively different ${\mathcal P}_z$--behaviour of two Stokes parameters: while $P_2$ is proportional to the degree of polarization of the ion beam, $P_1$ is unaffected by it. This behaviour is illustrated in Fig.~\ref{Fig2} where we display the Stokes parameters $P_1$ and $P_2$ as a function of the photon emission angle $\theta$ and for the recombination of polarized hydrogen--like uranium ions U$^{91+}$ with 100~keV Bessel electrons. Calculations have been performed for various ratios of the transverse to longitudinal momentum of Bessel electrons, which are conveniently parameterized by the opening angle $\tan \theta_p = \varkappa/p_z$, and for three degrees of ion polarization, ${\mathcal P}_z = 0$ (green solid line),  ${\mathcal P}_z = 0.5$ (blue dashed line) and ${\mathcal P}_z = 1.0$ (red dash--dotted line). For the case of unpolarized ion beam, ${\mathcal P}_z = 0$, the calculations clearly indicate that the Stokes parameter $P_2$ is identically $zero$ for all emission angles $\theta$, while the parameter $P_1$ reaches large positive values for $0 < \theta < 180$~deg. The situation changes qualitatively if electrons are recombined with longitudinally polarized ions. Namely, as can be seen from the figure, the (absolute value of the) second Stokes parameter increases linearly with ${\mathcal P}_z$, while the first parameter $P_1$ remains insensitive to ${\mathcal P}_z$. As we mentioned already above, non--zero values of the $P_2$ parameter indicate the rotation of the linear polarization of $K$--RR photons out of the reaction plane. Most naturally, this rotation can be parameterized by the tilt angle (\ref{eq:tilt_angle}) of the polarization ellipse. 

%
%
\begin{figure*}[t]
	\includegraphics[width=0.95\linewidth]{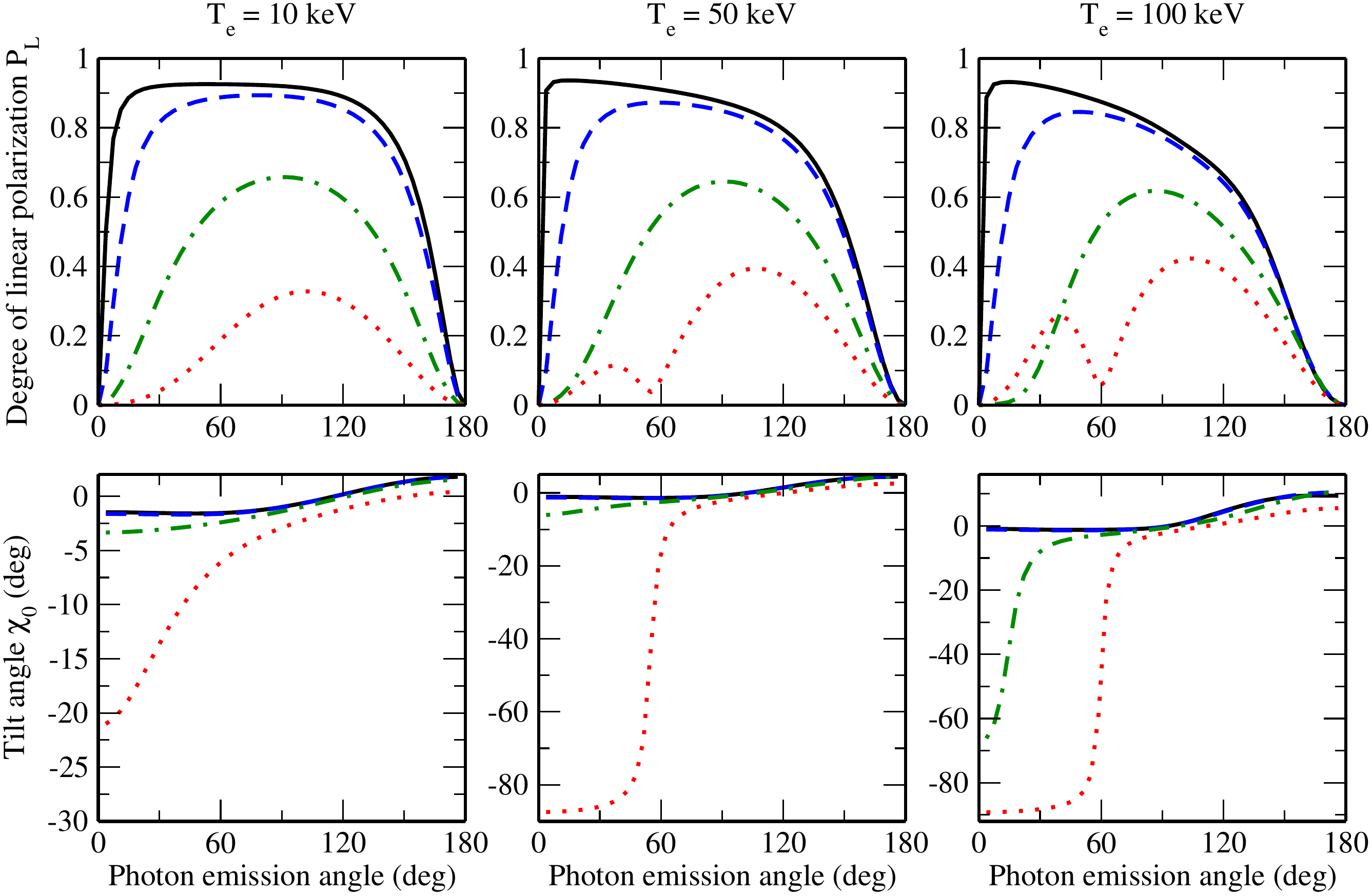}
	\caption{The degree of linear polarization $P_L = \sqrt{P_1^2 + P_2^2}$ (upper panel) and the polarization tilt angle $\chi_0$ (lower panel) of the photons, emitted in the $K$--RR of completely polarized, ${\mathcal P}_z = 1$, hydrogen--like ions. Calculations have been performed both for plane--wave (black solid line) and for twisted electrons with the opening angle $\theta_p$~=~10~deg (blue dashed line), $\theta_p$~=~30~deg (green dash--dotted line) and $\theta_p$~=~45~deg (red dotted line). The polarization parameters are obtained, moreover, for three electron kinetic energies: $T_e$~=~10~keV (left column), 50~keV (middle column) and 100~keV (right column). The results are presented in the ion rest frame.}
    \label{Fig3}
\end{figure*}
%
%

The polarization tilt angle $\chi_0$ for the radiative recombination of Bessel electrons with hydrogen--like ions can be obtained upon inserting $b$--averaged density matrix (\ref{eq:b-averaged_density_matrix_unstructured}) into Eqs.~(\ref{eq:Stokes_parameters_final_P1}), (\ref{eq:Stokes_parameters_final_P2}) and (\ref{eq:tilt_angle}). After some simple algebra one obtains:
\begin{equation}
    \label{eq:tilt_angle_unstructured}
    \tan 2\chi_0 = \frac{P_2}{P_1} = {\mathcal P}_z \, {\mathcal R}^{\rm (tw)}\left(Z, T_e, \theta_p ; \, \theta \right) \, ,
\end{equation}
where the function ${\mathcal R}^{\rm (tw)}\left(Z, T_e, \theta_p ; \, \theta \right)$ depends on the electron kinetic energy $T_e$, the nuclear charge $Z$ and the photon emission angle $\theta$. Moreover, in contrast to the plane--wave case, ${\mathcal R}^{\rm (tw)}$ can be modified also by varying the opening angle $\theta_p$ of the twisted electrons. It can be expected from the fact that both Stokes parameters $P_1$ and $P_2$ and, hence, their ratio depend on $\theta_p$, as can be seen from Fig.~\ref{Fig2}. This $\theta_p$--dependence can be used to enlarge the absolute values of the function ${\mathcal R}^{\rm (tw)}$ and, hence, to \textit{increase} the sensitivity of the tilt angle $\chi_0$ to the polarization of ion beam. In order to illustrate how twisted electrons may help in the spin--diagnostics of ions beams, we display in the lower panel of Fig.~\ref{Fig3} the tilt angle $\chi_0$ of the $K$--RR photons emitted in collisions of completely polarized U$^{91+}$ ions with plane--wave (black solid line) and Bessel electrons. For the latter case, calculations have been carried out for the opening angles $\theta_p$~=~10~deg (blue dashed line), 30~deg (green dash--dotted line) and 45~deg (red dotted line). We have investigated, moreover, the energy dependence of $\chi_0$ and present results for $T_e$~=~10~keV (left column), 50~keV (middle column) and 100~keV (right column). As seen from the figure, the tilt angle of the linear polarization of $K$--RR photons is rather sensitive to the state (plane--wave or Bessel) in which electrons are initially prepared. The effect becomes more pronounced for the forward photon emission, where $\chi_0$ can be strongly enhanced if twisted electrons with large opening angles $\theta_p$ are captured into the $K$--shell of hydrogen--like ions. For example, the tilt angle $\chi_0$, measured in the range $10 < \theta < 30$~deg, is increased by more than an order of magnitude if Bessel electrons with $\theta_p$~=~45~deg are used in place of the ``usual'' plane--wave ones. One can note that this increase of $\chi_0$ for larger opening angles $\theta_p$ of twisted light is associated with the reduction of the degree of linear polarization $P_L$, see upper panel of Fig.~\ref{Fig3}. Even being reduced, however, the $P_L$ reaches the values 10--40~\% for the forward photon emission $\theta < $~60~deg which makes the polarization measurements highly feasible. As mentioned already above, these measurements, when performed with the help of segmented Compton detectors, can accurately determine both $P_L$ and $\chi_0$. With the help of Bessel electrons, which allow \textit{controllable} enhancement of $\chi_0$, this Compton polarimetry can be used for the spin--diagnostics of ion beams even with small degrees of polarization ${\mathcal P}_z$.

%
%
\begin{figure*}[t]
	\includegraphics[width=0.95\linewidth]{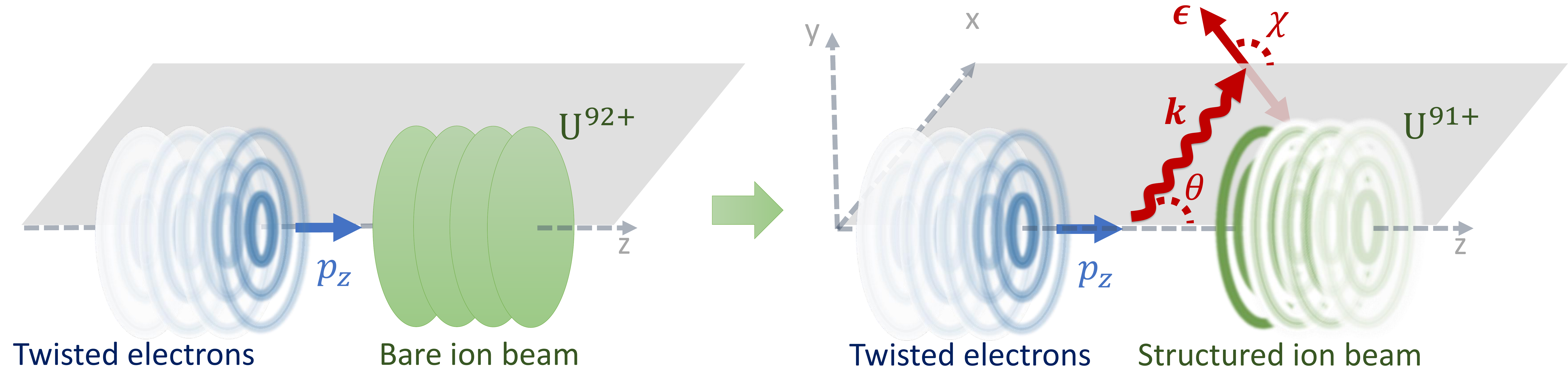}
	\caption{The geometry of the \textit{sequential} two--electron radiative recombination of initially bare (finally helium--like) uranium ions. In the first step, displayed in the left panel of the figure, the capture of Bessel electrons into the 1s$_{1/2}$ state of initially bare ions leads to formation of the structured U$^{91+}$ beam. A non--homogeneous intensity profile and vortex--like spin pattern of this beam affect the linear polarization of $K$--RR photons, emitted in collisions with the second (``probe'') twisted electron bunch, see right panel.}
    \label{Fig4}
\end{figure*}
%
%
%
%
\subsection{Recombination of structured ion beams}
\label{subsec:structured_beams}

In the previous section we have discussed the $K$--shell radiative recombination for collisions of Bessel electrons with the beams of hydrogen--like ions, whose sublevel population is independent of the position ${\bm b}$ of a particular ion. For such \textit{unstructured} beams the linear polarization of emitted photons, obtained upon integration over ${\bm b}$, reflects only the kinematic properties of Bessel electrons, but is insensitive to their TAM projection $m_j$. As seen from Eqs.~(\ref{eq:b-averaged_density_matrix}) and (\ref{eq:Fourier_transform}), the $m_j$--dependence of the $K$--RR linear polarization would indicate that the intensity and/or the polarization of the ion beam vary in its transverse cross--sectional plane. Thus, the radiative recombination of Bessel electrons can also serve as a tool for investigating the \textit{structured} ions beams. 

Of course, the sensitivity of the $K$--RR linear polarization to the internal beam structure depends on a shape and spin pattern of a particular beam. During the recent years, several proposals have been made to produce such structured ion beams by using, for example, immersed cathode technique \cite{FlK20} and the radiative capture of Bessel electrons \cite{MaP20}. The beams of hydrogen--like ions, obtained by the second (recombination) method, exhibit multiple--ring intensity profiles and vortex--like spin patterns that are described by the statistical tensors:
\begin{equation}
    \label{eq:rhokq_structured_beam}
    \rho_{kq}({\bm b}) = {\rm e}^{i q \phi_b} \, \rho_{kq}(b) \, ,
\end{equation}
with the impact parameter ${\bm b} = \left(b \cos\phi_b,\, b\sin\phi_b, \, 0 \right)$ defined in the transverse plane \cite{MaP20}. Below we will employ these beams as a ``testbed'' and will discuss their $K$--shell recombination with twisted electrons. If the ions were in the $1s_{1/2}$ hydrogenic state \textit{before} the recombination, one can use Eq.~(\ref{eq:rhokq_structured_beam}) with $k$~=~0, 1 and $q = -k, ..., k$ as well as formulas from Sec.~\ref{sec:theory} to calculate the Stokes parameters of the $K$--RR photons. Similar to before, the calculations have to be started with the evaluation of the Fourier transform:
\begin{eqnarray}
    \label{eq:Fourier_transform_structured}
    {\mathcal F}_{kq}({\bm p} - {\bm p}') &=& 2 \pi i^q \, {\rm e}^{ -i q \phi_Q} \nonumber \\
    &\times& \int{{\rm d}b \, b\, \rho_{kq}(b) \, J_q(Qb)} \ ,
\end{eqnarray}
where we assumed that the axes of the (electron and ion) beams coincide and introduced notation 
${\bm Q} = {\bm p}_\perp - {\bm p}'_\perp$. In contrast to the recombination of unstructured hydrogen--like ions, see Eq.~(\ref{eq:Fourier_transform_unstructured_beam}), this expression does not require equality of the transverse momenta ${\bm p}_\perp$ and ${\bm p}'_\perp$. This, in turn, implies that the density matrix (\ref{eq:b-averaged_density_matrix}) and, hence, the linear polarization of emitted photons can generally depend on the TAM projection $m_j$ of Bessel electrons even after integration over the impact parameter $b$.

%
%
\begin{figure*}[t]
	\includegraphics[width=0.95\linewidth]{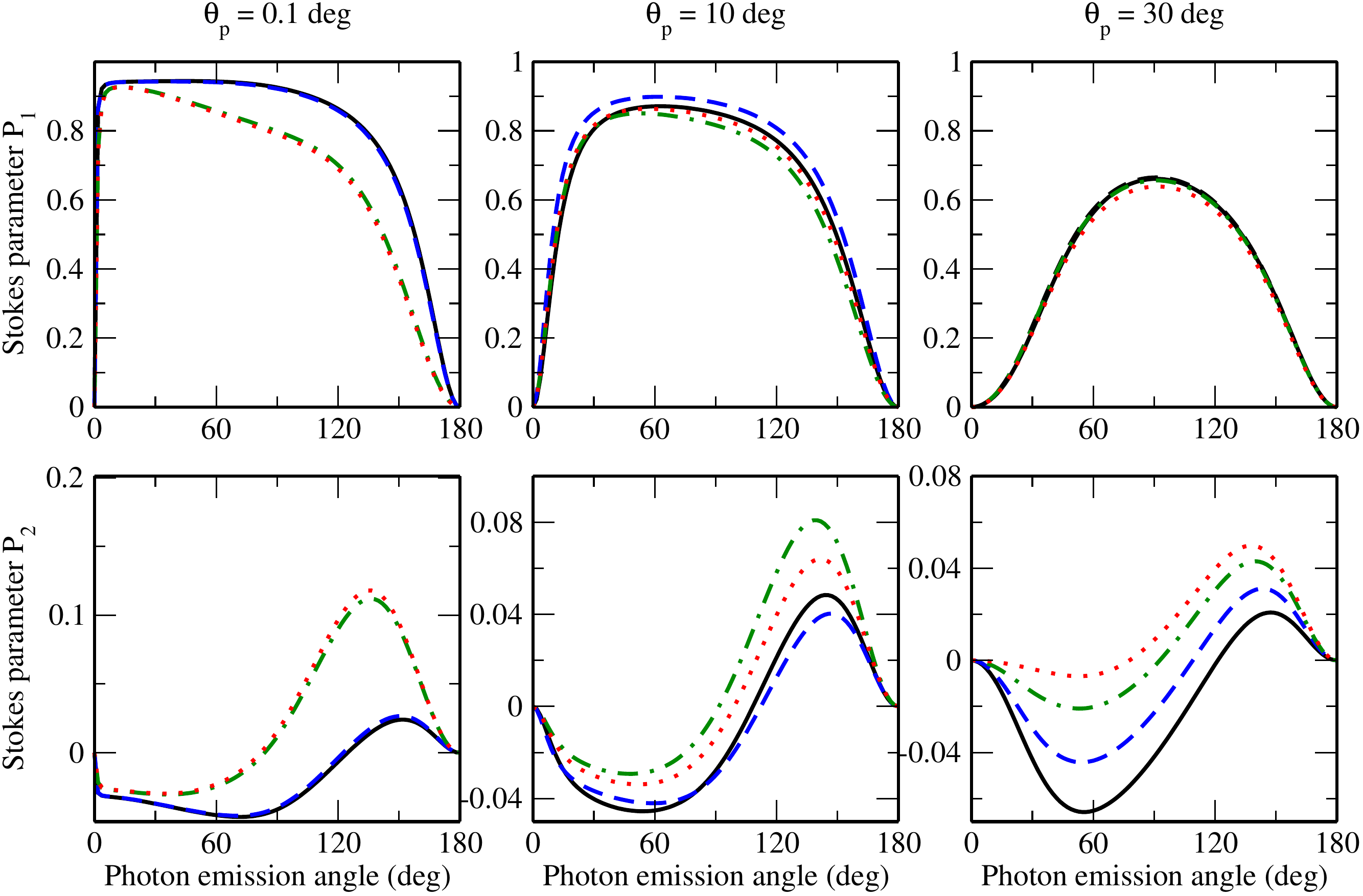}
	\caption{The Stokes parameters $P_1$ (upper panel) and $P_2$ (lower panel) of the $K$--RR photons, emitted in collisions of a structured beam of hydrogen--like uranium U$^{91+}$ ions with spin--unpolarized Bessel electrons carrying, however, well-defined projection of the total angular momentum $m_j = -3/2$ (black solid line), $m_j = -1/2$ (blue dashed line), $m_j = +1/2$ (green dash--dotted line) and $m_j = +3/2$ (red dotted line). Calculations have been performed for the electrons with kinetic energy 50~keV and opening angles $\theta_p$~=~0.1~deg (left column), $\theta_p$~=~10~deg (middle column) and $\theta_p$~=~30~deg (right column). The results are presented in the ion rest frame and for the case when the axes of both (electron and ion) beams coincide with each other.}
    \label{Fig5}
\end{figure*}
%
%

In order to investigate the $m_j$--dependence of the $K$--RR linear polarization we consider the scenario of \textit{sequential} radiative recombination of initially bare uranium ions U$^{92+}$ colliding with two spatially separated bunches of Bessel electrons, see Fig.~\ref{Fig4}. During collisions with electrons from the first bunch, which have TAM projection $m_j$~=~+1/2 and helicity $m_s$~=~+1/2 as well as kinetic energy 50~keV and opening angle $\theta_p$~=~15~deg in the ion frame, the hydrogen--like ions U$^{91+}$ in the ground $1s_{1/2}$ state can be produced. The intensity distribution and magnetic sublevel population of these U$^{91+}$ ions is described by Eq.~(\ref{eq:rhokq_structured_beam}) and can later affect the linear polarization of photons, emitted in the $K$--shell capture of electrons from the \textit{second} bunch. In Fig.~\ref{Fig5}, for example, we display the Stokes parameters $P_1$ and $P_2$ of such $K$--RR photons. Here, the calculations have been carried out for the recombination of (second, ``probe'') Bessel electrons with kinetic energy 50~keV, opening angles $\theta_p$~=~0.1~deg (left column), 10~deg (middle column) and 30~deg (right column), as well as with the TAM projections $m_j$~=~-3/2 (black solid line), -1/2 (blue dashed line), +1/2 (green dash--dotted line) and +3/2 (red dotted line). As seen from the figure, the linear polarization of K--$RR$ photons is very sensitive to the TAM projection of the ``probe'' electrons, and this $m_j$--sensitivity varies with the opening angle $\theta_p$. For example, the Stokes parameters depend mainly on the \textit{sign} of $m_j$ and are not very sensitive to its absolute value if $\theta_p$~=~0.1~deg. With the increase of the opening angle, the linear polarization of $K$--RR photons allows to distinguish both the sign and the absolute value of the TAM projection; the effect that is most pronounced for the parameter $P_2$. For instance, the second Stokes parameters, measured at the photon emission angle $\theta$~=~60~deg, increases from $P_2$~=~-0.065 for $m_j$~=~-3/2 to $P_2$~=~-0.006 for $m_j$~=~+3/2 if Bessel electrons have an opening angle $\theta_p$~=~30~deg. Such a remarkable enhancement of $P_2$ can be easily observed by modern Compton polarimeters and can help in the investigations of the intensity and spin patterns of structured ions beams.   

\begin{figure*}[t]
	\includegraphics[width=0.95\linewidth]{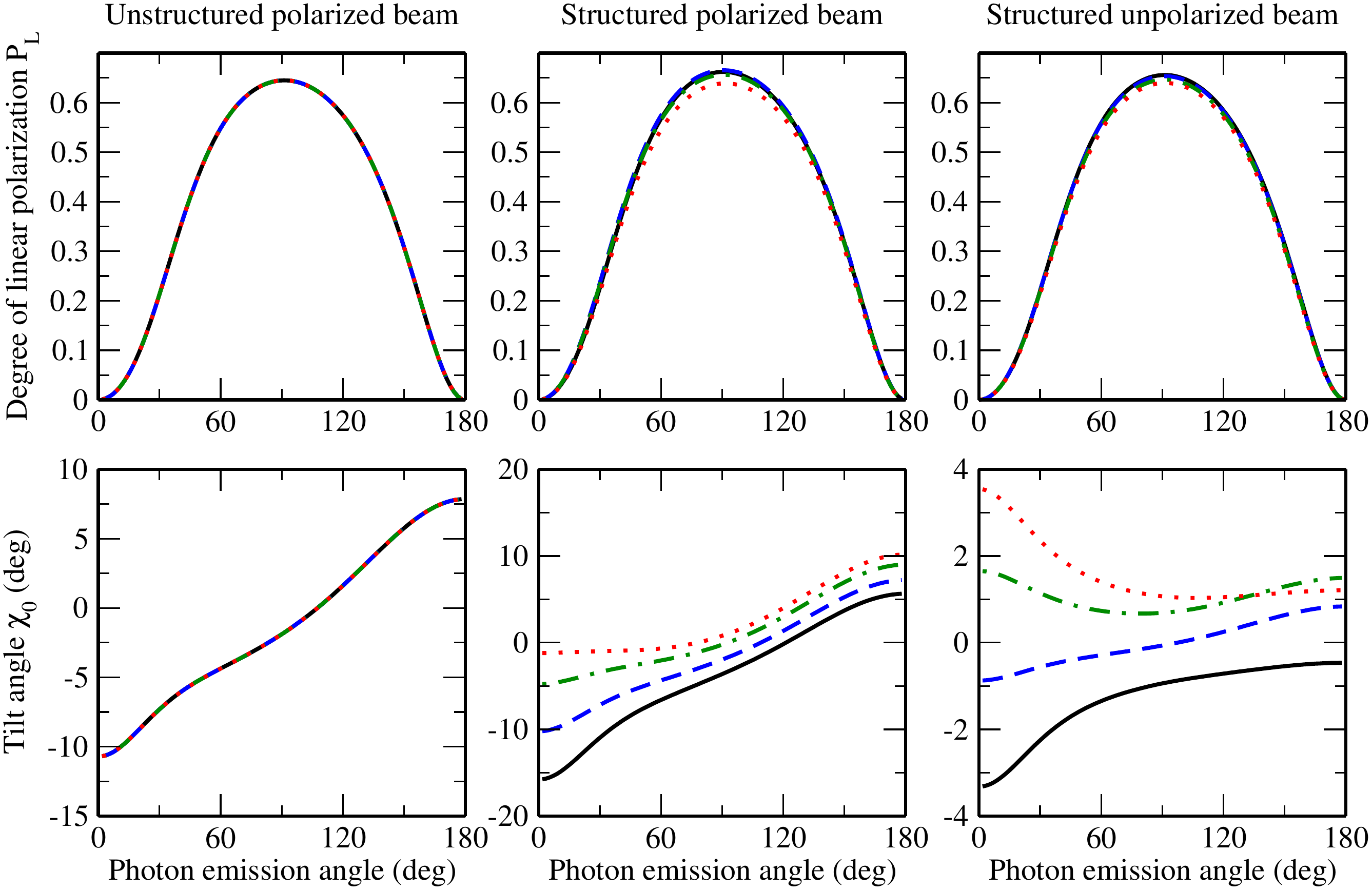}
	\caption{The degree of linear polarization $P_L = \sqrt{P_1^2 + P_2^2}$ (upper panel) and the polarization tilt angle $\chi_0$ (lower panel) of the photons, emitted in the $K$--RR of unstructured polarized (left column), structured polarized (middle panel) and structured unpolarized (right column) beams of hydrogen--like uranium ions. Calculations have been performed for unpolarized twisted electrons with the kinetic energy 50~keV and opening angle $\theta_p$~=~30~deg as well as with TAM projections $m_j = -3/2$ (black solid line), $m_j = -1/2$ (blue dashed line), $m_j = +1/2$ (green dash--dotted line) and $m_j = +3/2$ (red dotted line). The results are presented in the ion rest frame.}
    \label{Fig6}
\end{figure*}

As discussed already above, it is often more convenient to describe the linear polarization of light in terms of the polarization ellipse instead of two Stokes parameters $P_1$ and $P_2$. From Eq.~(\ref{eq:tilt_angle}) and Fig.~\ref{Fig5} one can expect that the tilt angle $\chi_0$ of this ellipse can also be used as a valuable tool for the diagnostics of structured ion beams. In Fig.~\ref{Fig6} we present both $\chi_0$ (lower panel) and the degree of polarization $P_L$ (upper panel) of x--rays, emitted in the $K$--RR of hydrogen--like uranium ions U$^{91+}$ with Bessel electrons. In contrast to results from Fig.~\ref{Fig5}, here we fix the opening angle $\theta_p$~=~30~deg of the ``probe'' electron beam but consider three different beams of hydrogen--like U$^{91+}$ ions. These beams were preliminary produced by the $K$--shell recombination of bare ions with polarized plane--wave electrons (left column) as well as with polarized (middle column) and unpolarized (right column) Bessel electrons, carrying TAM projection $m_j$~=~1/2 and opening angle $\theta_p$~=~15~deg. These three beams can referred to as (i) polarized unstructured, (ii) polarized structured and (iii) unpolarized structured beams, respectively. As seen from Fig.~\ref{Fig6}, the polarization--resolved measurements of $K$--RR photons allow to distinguish between these three cases. For example, as we know already from Sec.~\ref{subsec:nonstructured_beams} the parameters $P_L$ and $\chi_0$ of the polarization ellipse are insensitive to the TAM projection of the ``probing'' electrons colliding with an unstructured beam, cf. left panel of Fig.~\ref{Fig6}. In contrast, the tilt angle $\chi_0$ can vary significantly with $m_j$ if Bessel electrons recombine with structured (polarized and unpolarized) ion beams. This $m_j$--dependence is more pronounced for the $K$--RR of a polarized structured beam where the linear polarization of recombination photons, detected under small emission angles, can be tilted by more than 10~degrees if TAM projection of the ``probe'' electron beam is flipped from +3/2 to -3/2. Again, such a remarkable polarization rotation can be easily observed in present storage--ring experiments and can be used for the diagnostics of structured ion beams. 

Of course, the sensitivity of the $K$--RR linear polarization to the parameters of the ``probe'' Bessel electrons will strongly depend on internal structure of each particular (structured) ion beam and can differ a lot from that of our ``toy example'', displayed in Figs.~\ref{Fig5}--\ref{Fig6}. However, as follows from the \textit{general} theoretical approach, laid out in this work, the $m_j$--dependence of the polarization tilt angle $\chi_0$ will clearly indicate inhomogeneity of intensity profile and/or spin pattern of ion beams. The details about such (intensity and spin) structures can be obtained by combining experimental $K$--RR polarization data and theoretical predictions, based on Eqs.~(\ref{eq:Stokes_parameters_final_P1})--(\ref{eq:Fourier_transform}).  

%
%
%
%

%
%
\section{Summary and outlook}
\label{sec:summary}

In summary, we lay out a theoretical approach for the description of linear polarization of photons, emitted in the  radiative recombination of Bessel electrons with initially hydrogen--like (finally helium--like) heavy ions. A particular interest to the recombination arises from the fact that it can be used as a \textit{probe} process for the diagnostics of spin--polarization of ion beams in storage rings. In order to investigate whether twisted electrons are more preferable for such beam diagnostics then their plane--wave counterparts, we applied the density matrix theory and derived the polarization Stokes parameters of $K$--RR photons. These parameters were obtained for a realistic experimental scenario of collisions between \textit{macroscopic} electron and ion beams, and account both for characteristics of Bessel electrons and for the polarization of ions. While the derived expressions are general, we applied them in two case studies that deal with (spatially) unstructured and structured beams of hydrogen--like uranium U$^{91+}$ ions. For the first---unstructured---case we have re--visited the previously elaborated method of the spin diagnostics of ion beams that is based on the rotation of the linear polarization of $K$--RR photons out of the reaction plane. Results of our calculations indicate that the sensitivity of this method can be significantly enhanced if Bessel electrons instead of plane--wave ones are used in an experiment. In the second scenario we considered the capture of twisted electrons into the ground state of hydrogen--like U$^{91+}$ ions, whose local intensity and magnetic sublevel population vary depending on their position in a beam. Also for this case the linear polarization of $K$--RR photons provides a valuable tool for the investigation of complex internal structure of ion beams.    

The two case studies, presented in this work, indicate that the linear polarization of x--rays, emitted in the $K$--shell capture of Bessel electrons, can provide valuable information about (spatially--resolved) intensity and spin structure of heavy ion beams. In the future, therefore, the proposed method might become an effective diagnostic tool for ion storage ring experiments. It can be applied, in particular, to investigate how the polarization of an ion beam, produced by a standard optical pumping technique, is modified by the fields of dipole and quadrupole magnets along the circumference of a storage ring. The understanding and control of such a ``polarization dynamics'' is essential for experiments, aiming to explore atomic parity--violation phenomena in highly--charged ions. Based on the results of the present work, we currently develop the theoretical foundations of the polarization diagnostics of ions passing through magnetic fields. This study requires ion beam optics simulations and is currently underway. 

Yet another important effect, that has to be taken into account for the analysis of the (future) RR experiments with twisted electrons, is the production and focusing of electron beams. The use of the finite--size apertures may lead to a deviation of the realistic beams from the simple Bessel solutions, given by Eq.~(\ref{eq:wave_function_Bessel}); the effect which becomes more pronounced as the distance from the beam axis increases. In turn, it can lead to the modification of the linear polarization of RR photons, which will reflect now also the way of production of twisted electrons. The theoretical analysis of the electron beam focusing effects requires application of the wave--packet approach, which has been developed recently for the Bessel and Airy solutions \cite{KaK17,Kar15}. The further extension of this approach towards relativistic electron--ion collisions at storage rings will be the subject of our forthcoming publications.

\section{Acknowledgements}
\label{sec:acknowledgements}

Funded by the Russian Science Foundation (Grant No. 20-62-46006). A.~A.~P. acknowledges support from the Deutsche Forschungsgemeinschaft (DFG, German Research Foundation) under Germany's Excellence Strategy -- EXC-2123 QuantumFrontiers -- 390837967. 

%
%
%
%
\bibliography{./b20.twisted_electrons_H-like}

\begin{thebibliography}{38}%
\makeatletter
\providecommand \@ifxundefined [1]{%
 \@ifx{#1\undefined}
}%
\providecommand \@ifnum [1]{%
 \ifnum #1\expandafter \@firstoftwo
 \else \expandafter \@secondoftwo
 \fi
}%
\providecommand \@ifx [1]{%
 \ifx #1\expandafter \@firstoftwo
 \else \expandafter \@secondoftwo
 \fi
}%
\providecommand \natexlab [1]{#1}%
\providecommand \enquote  [1]{``#1''}%
\providecommand \bibnamefont  [1]{#1}%
\providecommand \bibfnamefont [1]{#1}%
\providecommand \citenamefont [1]{#1}%
\providecommand \href@noop [0]{\@secondoftwo}%
\providecommand \href [0]{\begingroup \@sanitize@url \@href}%
\providecommand \@href[1]{\@@startlink{#1}\@@href}%
\providecommand \@@href[1]{\endgroup#1\@@endlink}%
\providecommand \@sanitize@url [0]{\catcode `\\12\catcode `\$12\catcode
  `\&12\catcode `\#12\catcode `\^12\catcode `\_12\catcode `\%12\relax}%
\providecommand \@@startlink[1]{}%
\providecommand \@@endlink[0]{}%
\providecommand \url  [0]{\begingroup\@sanitize@url \@url }%
\providecommand \@url [1]{\endgroup\@href {#1}{\urlprefix }}%
\providecommand \urlprefix  [0]{URL }%
\providecommand \Eprint [0]{\href }%
\providecommand \doibase [0]{https://doi.org/}%
\providecommand \selectlanguage [0]{\@gobble}%
\providecommand \bibinfo  [0]{\@secondoftwo}%
\providecommand \bibfield  [0]{\@secondoftwo}%
\providecommand \translation [1]{[#1]}%
\providecommand \BibitemOpen [0]{}%
\providecommand \bibitemStop [0]{}%
\providecommand \bibitemNoStop [0]{.\EOS\space}%
\providecommand \EOS [0]{\spacefactor3000\relax}%
\providecommand \BibitemShut  [1]{\csname bibitem#1\endcsname}%
\let\auto@bib@innerbib\@empty
\bibitem [{\citenamefont {Pajek}\ and\ \citenamefont {Schuch}(1992)}]{PaS92}%
  \BibitemOpen
  \bibfield  {author} {\bibinfo {author} {\bibfnamefont {M.}~\bibnamefont
  {Pajek}}\ and\ \bibinfo {author} {\bibfnamefont {R.}~\bibnamefont {Schuch}},\
  }\bibfield  {title} {\bibinfo {title} {Radiative recombination of bare ions
  with low-energy free electrons},\ }\href
  {https://doi.org/10.1103/PhysRevA.45.7894} {\bibfield  {journal} {\bibinfo
  {journal} {Phys. Rev. A}\ }\textbf {\bibinfo {volume} {45}},\ \bibinfo
  {pages} {7894} (\bibinfo {year} {1992})}\BibitemShut {NoStop}%
\bibitem [{\citenamefont {Eichler}\ and\ \citenamefont
  {Meyerhof}(1995)}]{EiM95}%
  \BibitemOpen
  \bibfield  {author} {\bibinfo {author} {\bibfnamefont {J.}~\bibnamefont
  {Eichler}}\ and\ \bibinfo {author} {\bibfnamefont {W.}~\bibnamefont
  {Meyerhof}},\ }\href {https://books.google.de/books?id=RNTvAAAAMAAJ} {\emph
  {\bibinfo {title} {Relativistic Atomic Collisions}}}\ (\bibinfo  {publisher}
  {Academic Press},\ \bibinfo {year} {1995})\BibitemShut {NoStop}%
\bibitem [{\citenamefont {Fritzsche}\ \emph {et~al.}(2005)\citenamefont
  {Fritzsche}, \citenamefont {Surzhykov},\ and\ \citenamefont
  {St\"ohlker}}]{FrS05}%
  \BibitemOpen
  \bibfield  {author} {\bibinfo {author} {\bibfnamefont {S.}~\bibnamefont
  {Fritzsche}}, \bibinfo {author} {\bibfnamefont {A.}~\bibnamefont
  {Surzhykov}},\ and\ \bibinfo {author} {\bibfnamefont {T.}~\bibnamefont
  {St\"ohlker}},\ }\bibfield  {title} {\bibinfo {title} {Radiative
  recombination into high-{Z} few-electron ions: {C}ross sections and angular
  distributions},\ }\href {https://doi.org/10.1103/PhysRevA.72.012704}
  {\bibfield  {journal} {\bibinfo  {journal} {Phys. Rev. A}\ }\textbf {\bibinfo
  {volume} {72}},\ \bibinfo {pages} {012704} (\bibinfo {year}
  {2005})}\BibitemShut {NoStop}%
\bibitem [{\citenamefont {{J.~Eichler and Th.~St\"ohlker}}(2007)}]{EiS07}%
  \BibitemOpen
  \bibfield  {author} {\bibinfo {author} {\bibnamefont {{J.~Eichler and
  Th.~St\"ohlker}}},\ }\bibfield  {title} {\bibinfo {title} {{Radiative
  electron capture in relativistic ion--atom collisions and the photoelectric
  effect in hydrogen--like high--{Z} systems}},\ }\href
  {https://doi.org/https://doi.org/10.1016/j.physrep.2006.11.003} {\bibfield
  {journal} {\bibinfo  {journal} {Physics Reports}\ }\textbf {\bibinfo {volume}
  {439}},\ \bibinfo {pages} {1 } (\bibinfo {year} {2007})}\BibitemShut
  {NoStop}%
\bibitem [{\citenamefont {Surzhykov}\ \emph {et~al.}(2006)\citenamefont
  {Surzhykov}, \citenamefont {Jentschura}, \citenamefont {St\"ohlker},\ and\
  \citenamefont {Fritzsche}}]{SuJ06}%
  \BibitemOpen
  \bibfield  {author} {\bibinfo {author} {\bibfnamefont {A.}~\bibnamefont
  {Surzhykov}}, \bibinfo {author} {\bibfnamefont {U.~D.}\ \bibnamefont
  {Jentschura}}, \bibinfo {author} {\bibfnamefont {T.}~\bibnamefont
  {St\"ohlker}},\ and\ \bibinfo {author} {\bibfnamefont {S.}~\bibnamefont
  {Fritzsche}},\ }\bibfield  {title} {\bibinfo {title} {Radiative electron
  capture into high-{Z} few-electron ions: {A}lignment of the excited ionic
  states},\ }\href {https://doi.org/10.1103/PhysRevA.73.032716} {\bibfield
  {journal} {\bibinfo  {journal} {Phys. Rev. A}\ }\textbf {\bibinfo {volume}
  {73}},\ \bibinfo {pages} {032716} (\bibinfo {year} {2006})}\BibitemShut
  {NoStop}%
\bibitem [{\citenamefont {St\"ohlker}\ \emph {et~al.}(1999)\citenamefont
  {St\"ohlker}, \citenamefont {Ludziejewski}, \citenamefont {Bosch},
  \citenamefont {Dunford}, \citenamefont {Kozhuharov}, \citenamefont {Mokler},
  \citenamefont {Beyer}, \citenamefont {Brinzanescu}, \citenamefont {Franzke},
  \citenamefont {Eichler}, \citenamefont {Griegal}, \citenamefont {Hagmann},
  \citenamefont {Ichihara}, \citenamefont {Kr\"amer}, \citenamefont {Lekki},
  \citenamefont {Liesen}, \citenamefont {Nolden}, \citenamefont {Reich},
  \citenamefont {Rymuza}, \citenamefont {Stachura}, \citenamefont {Steck},
  \citenamefont {Swiat},\ and\ \citenamefont {Warczak}}]{StL99}%
  \BibitemOpen
  \bibfield  {author} {\bibinfo {author} {\bibfnamefont {T.}~\bibnamefont
  {St\"ohlker}}, \bibinfo {author} {\bibfnamefont {T.}~\bibnamefont
  {Ludziejewski}}, \bibinfo {author} {\bibfnamefont {F.}~\bibnamefont {Bosch}},
  \bibinfo {author} {\bibfnamefont {R.~W.}\ \bibnamefont {Dunford}}, \bibinfo
  {author} {\bibfnamefont {C.}~\bibnamefont {Kozhuharov}}, \bibinfo {author}
  {\bibfnamefont {P.~H.}\ \bibnamefont {Mokler}}, \bibinfo {author}
  {\bibfnamefont {H.~F.}\ \bibnamefont {Beyer}}, \bibinfo {author}
  {\bibfnamefont {O.}~\bibnamefont {Brinzanescu}}, \bibinfo {author}
  {\bibfnamefont {B.}~\bibnamefont {Franzke}}, \bibinfo {author} {\bibfnamefont
  {J.}~\bibnamefont {Eichler}}, \bibinfo {author} {\bibfnamefont
  {A.}~\bibnamefont {Griegal}}, \bibinfo {author} {\bibfnamefont
  {S.}~\bibnamefont {Hagmann}}, \bibinfo {author} {\bibfnamefont
  {A.}~\bibnamefont {Ichihara}}, \bibinfo {author} {\bibfnamefont
  {A.}~\bibnamefont {Kr\"amer}}, \bibinfo {author} {\bibfnamefont
  {J.}~\bibnamefont {Lekki}}, \bibinfo {author} {\bibfnamefont
  {D.}~\bibnamefont {Liesen}}, \bibinfo {author} {\bibfnamefont
  {F.}~\bibnamefont {Nolden}}, \bibinfo {author} {\bibfnamefont
  {H.}~\bibnamefont {Reich}}, \bibinfo {author} {\bibfnamefont
  {P.}~\bibnamefont {Rymuza}}, \bibinfo {author} {\bibfnamefont
  {Z.}~\bibnamefont {Stachura}}, \bibinfo {author} {\bibfnamefont
  {M.}~\bibnamefont {Steck}}, \bibinfo {author} {\bibfnamefont
  {P.}~\bibnamefont {Swiat}},\ and\ \bibinfo {author} {\bibfnamefont
  {A.}~\bibnamefont {Warczak}},\ }\bibfield  {title} {\bibinfo {title} {Angular
  distribution studies for the time-reversed photoionization process in
  hydrogenlike uranium: {T}he identification of spin-flip transitions},\ }\href
  {https://doi.org/10.1103/PhysRevLett.82.3232} {\bibfield  {journal} {\bibinfo
   {journal} {Phys. Rev. Lett.}\ }\textbf {\bibinfo {volume} {82}},\ \bibinfo
  {pages} {3232} (\bibinfo {year} {1999})}\BibitemShut {NoStop}%
\bibitem [{\citenamefont {St\"ohlker}\ \emph {et~al.}(2001)\citenamefont
  {St\"ohlker}, \citenamefont {Ma}, \citenamefont {Ludziejewski}, \citenamefont
  {Beyer}, \citenamefont {Bosch}, \citenamefont {Brinzanescu}, \citenamefont
  {Dunford}, \citenamefont {Eichler}, \citenamefont {Hagmann}, \citenamefont
  {Ichihara}, \citenamefont {Kozhuharov}, \citenamefont {Kr\"amer},
  \citenamefont {Liesen}, \citenamefont {Mokler}, \citenamefont {Stachura},
  \citenamefont {Swiat},\ and\ \citenamefont {Warczak}}]{StM01}%
  \BibitemOpen
  \bibfield  {author} {\bibinfo {author} {\bibfnamefont {T.}~\bibnamefont
  {St\"ohlker}}, \bibinfo {author} {\bibfnamefont {X.}~\bibnamefont {Ma}},
  \bibinfo {author} {\bibfnamefont {T.}~\bibnamefont {Ludziejewski}}, \bibinfo
  {author} {\bibfnamefont {H.~F.}\ \bibnamefont {Beyer}}, \bibinfo {author}
  {\bibfnamefont {F.}~\bibnamefont {Bosch}}, \bibinfo {author} {\bibfnamefont
  {O.}~\bibnamefont {Brinzanescu}}, \bibinfo {author} {\bibfnamefont {R.~W.}\
  \bibnamefont {Dunford}}, \bibinfo {author} {\bibfnamefont {J.}~\bibnamefont
  {Eichler}}, \bibinfo {author} {\bibfnamefont {S.}~\bibnamefont {Hagmann}},
  \bibinfo {author} {\bibfnamefont {A.}~\bibnamefont {Ichihara}}, \bibinfo
  {author} {\bibfnamefont {C.}~\bibnamefont {Kozhuharov}}, \bibinfo {author}
  {\bibfnamefont {A.}~\bibnamefont {Kr\"amer}}, \bibinfo {author}
  {\bibfnamefont {D.}~\bibnamefont {Liesen}}, \bibinfo {author} {\bibfnamefont
  {P.~H.}\ \bibnamefont {Mokler}}, \bibinfo {author} {\bibfnamefont
  {Z.}~\bibnamefont {Stachura}}, \bibinfo {author} {\bibfnamefont
  {P.}~\bibnamefont {Swiat}},\ and\ \bibinfo {author} {\bibfnamefont
  {A.}~\bibnamefont {Warczak}},\ }\bibfield  {title} {\bibinfo {title}
  {Near-threshold photoionization of hydrogenlike uranium studied in ion-atom
  collisions via the time-reversed process},\ }\href
  {https://doi.org/10.1103/PhysRevLett.86.983} {\bibfield  {journal} {\bibinfo
  {journal} {Phys. Rev. Lett.}\ }\textbf {\bibinfo {volume} {86}},\ \bibinfo
  {pages} {983} (\bibinfo {year} {2001})}\BibitemShut {NoStop}%
\bibitem [{\citenamefont {Tashenov}\ \emph {et~al.}(2006)\citenamefont
  {Tashenov}, \citenamefont {St\"ohlker}, \citenamefont
  {Bana\ifmmode~\acute{s}\else \'{s}\fi{}}, \citenamefont {Beckert},
  \citenamefont {Beller}, \citenamefont {Beyer}, \citenamefont {Bosch},
  \citenamefont {Fritzsche}, \citenamefont {Gumberidze}, \citenamefont
  {Hagmann}, \citenamefont {Kozhuharov}, \citenamefont {Krings}, \citenamefont
  {Liesen}, \citenamefont {Nolden}, \citenamefont {Protic}, \citenamefont
  {Sierpowski}, \citenamefont {Spillmann}, \citenamefont {Steck},\ and\
  \citenamefont {Surzhykov}}]{TaS06}%
  \BibitemOpen
  \bibfield  {author} {\bibinfo {author} {\bibfnamefont {S.}~\bibnamefont
  {Tashenov}}, \bibinfo {author} {\bibfnamefont {T.}~\bibnamefont
  {St\"ohlker}}, \bibinfo {author} {\bibfnamefont {D.}~\bibnamefont
  {Bana\ifmmode~\acute{s}\else \'{s}\fi{}}}, \bibinfo {author} {\bibfnamefont
  {K.}~\bibnamefont {Beckert}}, \bibinfo {author} {\bibfnamefont
  {P.}~\bibnamefont {Beller}}, \bibinfo {author} {\bibfnamefont {H.~F.}\
  \bibnamefont {Beyer}}, \bibinfo {author} {\bibfnamefont {F.}~\bibnamefont
  {Bosch}}, \bibinfo {author} {\bibfnamefont {S.}~\bibnamefont {Fritzsche}},
  \bibinfo {author} {\bibfnamefont {A.}~\bibnamefont {Gumberidze}}, \bibinfo
  {author} {\bibfnamefont {S.}~\bibnamefont {Hagmann}}, \bibinfo {author}
  {\bibfnamefont {C.}~\bibnamefont {Kozhuharov}}, \bibinfo {author}
  {\bibfnamefont {T.}~\bibnamefont {Krings}}, \bibinfo {author} {\bibfnamefont
  {D.}~\bibnamefont {Liesen}}, \bibinfo {author} {\bibfnamefont
  {F.}~\bibnamefont {Nolden}}, \bibinfo {author} {\bibfnamefont
  {D.}~\bibnamefont {Protic}}, \bibinfo {author} {\bibfnamefont
  {D.}~\bibnamefont {Sierpowski}}, \bibinfo {author} {\bibfnamefont
  {U.}~\bibnamefont {Spillmann}}, \bibinfo {author} {\bibfnamefont
  {M.}~\bibnamefont {Steck}},\ and\ \bibinfo {author} {\bibfnamefont
  {A.}~\bibnamefont {Surzhykov}},\ }\bibfield  {title} {\bibinfo {title} {First
  measurement of the linear polarization of radiative electron capture
  transitions},\ }\href {https://doi.org/10.1103/PhysRevLett.97.223202}
  {\bibfield  {journal} {\bibinfo  {journal} {Phys. Rev. Lett.}\ }\textbf
  {\bibinfo {volume} {97}},\ \bibinfo {pages} {223202} (\bibinfo {year}
  {2006})}\BibitemShut {NoStop}%
\bibitem [{\citenamefont {Vockert}\ \emph {et~al.}(2019)\citenamefont
  {Vockert}, \citenamefont {Weber}, \citenamefont {Br\"auning}, \citenamefont
  {Surzhykov}, \citenamefont {Brandau}, \citenamefont {Fritzsche},
  \citenamefont {Geyer}, \citenamefont {Hagmann}, \citenamefont {Hess},
  \citenamefont {Kozhuharov}, \citenamefont {M\"artin}, \citenamefont
  {Petridis}, \citenamefont {Hess}, \citenamefont {Trotsenko}, \citenamefont
  {Litvinov}, \citenamefont {Glorius}, \citenamefont {Gumberidze},
  \citenamefont {Steck}, \citenamefont {Litvinov}, \citenamefont {Ga\ss{}ner},
  \citenamefont {Hillenbrand}, \citenamefont {Lestinsky}, \citenamefont
  {Nolden}, \citenamefont {Sanjari}, \citenamefont {Popp}, \citenamefont
  {Trageser}, \citenamefont {Winters}, \citenamefont {Spillmann}, \citenamefont
  {Krings},\ and\ \citenamefont {St\"ohlker}}]{VoW19}%
  \BibitemOpen
  \bibfield  {author} {\bibinfo {author} {\bibfnamefont {M.}~\bibnamefont
  {Vockert}}, \bibinfo {author} {\bibfnamefont {G.}~\bibnamefont {Weber}},
  \bibinfo {author} {\bibfnamefont {H.}~\bibnamefont {Br\"auning}}, \bibinfo
  {author} {\bibfnamefont {A.}~\bibnamefont {Surzhykov}}, \bibinfo {author}
  {\bibfnamefont {C.}~\bibnamefont {Brandau}}, \bibinfo {author} {\bibfnamefont
  {S.}~\bibnamefont {Fritzsche}}, \bibinfo {author} {\bibfnamefont
  {S.}~\bibnamefont {Geyer}}, \bibinfo {author} {\bibfnamefont
  {S.}~\bibnamefont {Hagmann}}, \bibinfo {author} {\bibfnamefont
  {S.}~\bibnamefont {Hess}}, \bibinfo {author} {\bibfnamefont {C.}~\bibnamefont
  {Kozhuharov}}, \bibinfo {author} {\bibfnamefont {R.}~\bibnamefont
  {M\"artin}}, \bibinfo {author} {\bibfnamefont {N.}~\bibnamefont {Petridis}},
  \bibinfo {author} {\bibfnamefont {R.}~\bibnamefont {Hess}}, \bibinfo {author}
  {\bibfnamefont {S.}~\bibnamefont {Trotsenko}}, \bibinfo {author}
  {\bibfnamefont {Y.~A.}\ \bibnamefont {Litvinov}}, \bibinfo {author}
  {\bibfnamefont {J.}~\bibnamefont {Glorius}}, \bibinfo {author} {\bibfnamefont
  {A.}~\bibnamefont {Gumberidze}}, \bibinfo {author} {\bibfnamefont
  {M.}~\bibnamefont {Steck}}, \bibinfo {author} {\bibfnamefont
  {S.}~\bibnamefont {Litvinov}}, \bibinfo {author} {\bibfnamefont
  {T.}~\bibnamefont {Ga\ss{}ner}}, \bibinfo {author} {\bibfnamefont {P.-M.}\
  \bibnamefont {Hillenbrand}}, \bibinfo {author} {\bibfnamefont
  {M.}~\bibnamefont {Lestinsky}}, \bibinfo {author} {\bibfnamefont
  {F.}~\bibnamefont {Nolden}}, \bibinfo {author} {\bibfnamefont {M.~S.}\
  \bibnamefont {Sanjari}}, \bibinfo {author} {\bibfnamefont {U.}~\bibnamefont
  {Popp}}, \bibinfo {author} {\bibfnamefont {C.}~\bibnamefont {Trageser}},
  \bibinfo {author} {\bibfnamefont {D.~F.~A.}\ \bibnamefont {Winters}},
  \bibinfo {author} {\bibfnamefont {U.}~\bibnamefont {Spillmann}}, \bibinfo
  {author} {\bibfnamefont {T.}~\bibnamefont {Krings}},\ and\ \bibinfo {author}
  {\bibfnamefont {T.}~\bibnamefont {St\"ohlker}},\ }\bibfield  {title}
  {\bibinfo {title} {Radiative electron capture as a tunable source of highly
  linearly polarized x rays},\ }\href
  {https://doi.org/10.1103/PhysRevA.99.052702} {\bibfield  {journal} {\bibinfo
  {journal} {Phys. Rev. A}\ }\textbf {\bibinfo {volume} {99}},\ \bibinfo
  {pages} {052702} (\bibinfo {year} {2019})}\BibitemShut {NoStop}%
\bibitem [{\citenamefont {Spillmann}\ \emph {et~al.}(2008)\citenamefont
  {Spillmann}, \citenamefont {Br\"auning}, \citenamefont {Hess}, \citenamefont
  {Beyer}, \citenamefont {St\"ohlker}, \citenamefont {Dousse}, \citenamefont
  {Protic},\ and\ \citenamefont {Krings}}]{SpB08}%
  \BibitemOpen
  \bibfield  {author} {\bibinfo {author} {\bibfnamefont {U.}~\bibnamefont
  {Spillmann}}, \bibinfo {author} {\bibfnamefont {H.}~\bibnamefont
  {Br\"auning}}, \bibinfo {author} {\bibfnamefont {S.}~\bibnamefont {Hess}},
  \bibinfo {author} {\bibfnamefont {H.}~\bibnamefont {Beyer}}, \bibinfo
  {author} {\bibfnamefont {T.}~\bibnamefont {St\"ohlker}}, \bibinfo {author}
  {\bibfnamefont {J.-C.}\ \bibnamefont {Dousse}}, \bibinfo {author}
  {\bibfnamefont {D.}~\bibnamefont {Protic}},\ and\ \bibinfo {author}
  {\bibfnamefont {T.}~\bibnamefont {Krings}},\ }\bibfield  {title} {\bibinfo
  {title} {Performance of a {G}e-microstrip imaging detector and polarimeter},\
  }\href {https://doi.org/10.1063/1.2963046} {\bibfield  {journal} {\bibinfo
  {journal} {Review of Scientific Instruments}\ }\textbf {\bibinfo {volume}
  {79}},\ \bibinfo {pages} {083101} (\bibinfo {year} {2008})}\BibitemShut
  {NoStop}%
\bibitem [{\citenamefont {Vockert}\ \emph {et~al.}(2017)\citenamefont
  {Vockert}, \citenamefont {Weber}, \citenamefont {Spillmann}, \citenamefont
  {Krings}, \citenamefont {Herdrich},\ and\ \citenamefont
  {St\"ohlker}}]{VoW17}%
  \BibitemOpen
  \bibfield  {author} {\bibinfo {author} {\bibfnamefont {M.}~\bibnamefont
  {Vockert}}, \bibinfo {author} {\bibfnamefont {G.}~\bibnamefont {Weber}},
  \bibinfo {author} {\bibfnamefont {U.}~\bibnamefont {Spillmann}}, \bibinfo
  {author} {\bibfnamefont {T.}~\bibnamefont {Krings}}, \bibinfo {author}
  {\bibfnamefont {M.}~\bibnamefont {Herdrich}},\ and\ \bibinfo {author}
  {\bibfnamefont {T.}~\bibnamefont {St\"ohlker}},\ }\bibfield  {title}
  {\bibinfo {title} {Commissioning of a {S}i({L}i) {C}ompton polarimeter with
  improved energy resolution},\ }\href
  {https://doi.org/https://doi.org/10.1016/j.nimb.2017.05.035} {\bibfield
  {journal} {\bibinfo  {journal} {Nuclear Instruments and Methods in Physics
  Research Section B: Beam Interactions with Materials and Atoms}\ }\textbf
  {\bibinfo {volume} {408}},\ \bibinfo {pages} {313} (\bibinfo {year}
  {2017})}\BibitemShut {NoStop}%
\bibitem [{\citenamefont {Surzhykov}\ \emph {et~al.}(2011)\citenamefont
  {Surzhykov}, \citenamefont {Artemyev},\ and\ \citenamefont
  {Yerokhin}}]{SuA11}%
  \BibitemOpen
  \bibfield  {author} {\bibinfo {author} {\bibfnamefont {A.}~\bibnamefont
  {Surzhykov}}, \bibinfo {author} {\bibfnamefont {A.~N.}\ \bibnamefont
  {Artemyev}},\ and\ \bibinfo {author} {\bibfnamefont {V.~A.}\ \bibnamefont
  {Yerokhin}},\ }\bibfield  {title} {\bibinfo {title} {Interelectronic
  interaction effects on the polarization of recombination photons},\ }\href
  {https://doi.org/10.1103/PhysRevA.83.062710} {\bibfield  {journal} {\bibinfo
  {journal} {Phys. Rev. A}\ }\textbf {\bibinfo {volume} {83}},\ \bibinfo
  {pages} {062710} (\bibinfo {year} {2011})}\BibitemShut {NoStop}%
\bibitem [{\citenamefont {Holmberg}\ \emph {et~al.}(2015)\citenamefont
  {Holmberg}, \citenamefont {Artemyev}, \citenamefont {Surzhykov},
  \citenamefont {Yerokhin},\ and\ \citenamefont {St\"ohlker}}]{HoA15}%
  \BibitemOpen
  \bibfield  {author} {\bibinfo {author} {\bibfnamefont {J.}~\bibnamefont
  {Holmberg}}, \bibinfo {author} {\bibfnamefont {A.~N.}\ \bibnamefont
  {Artemyev}}, \bibinfo {author} {\bibfnamefont {A.}~\bibnamefont {Surzhykov}},
  \bibinfo {author} {\bibfnamefont {V.~A.}\ \bibnamefont {Yerokhin}},\ and\
  \bibinfo {author} {\bibfnamefont {T.}~\bibnamefont {St\"ohlker}},\ }\bibfield
   {title} {\bibinfo {title} {{QED} corrections to radiative recombination and
  radiative decay of heavy hydrogenlike ions},\ }\href
  {https://doi.org/10.1103/PhysRevA.92.042510} {\bibfield  {journal} {\bibinfo
  {journal} {Phys. Rev. A}\ }\textbf {\bibinfo {volume} {92}},\ \bibinfo
  {pages} {042510} (\bibinfo {year} {2015})}\BibitemShut {NoStop}%
\bibitem [{\citenamefont {Surzhykov}\ \emph {et~al.}(2005)\citenamefont
  {Surzhykov}, \citenamefont {Fritzsche}, \citenamefont {St\"ohlker},\ and\
  \citenamefont {Tashenov}}]{SuF05}%
  \BibitemOpen
  \bibfield  {author} {\bibinfo {author} {\bibfnamefont {A.}~\bibnamefont
  {Surzhykov}}, \bibinfo {author} {\bibfnamefont {S.}~\bibnamefont
  {Fritzsche}}, \bibinfo {author} {\bibfnamefont {T.}~\bibnamefont
  {St\"ohlker}},\ and\ \bibinfo {author} {\bibfnamefont {S.}~\bibnamefont
  {Tashenov}},\ }\bibfield  {title} {\bibinfo {title} {Application of radiative
  electron capture for the diagnostics of spin-polarized ion beams at storage
  rings},\ }\href {https://doi.org/10.1103/PhysRevLett.94.203202} {\bibfield
  {journal} {\bibinfo  {journal} {Phys. Rev. Lett.}\ }\textbf {\bibinfo
  {volume} {94}},\ \bibinfo {pages} {203202} (\bibinfo {year}
  {2005})}\BibitemShut {NoStop}%
\bibitem [{\citenamefont {Budker}\ \emph {et~al.}(2020)\citenamefont {Budker},
  \citenamefont {Crespo López-Urrutia}, \citenamefont {Derevianko},
  \citenamefont {Flambaum}, \citenamefont {Krasny}, \citenamefont {Petrenko},
  \citenamefont {Pustelny}, \citenamefont {Surzhykov}, \citenamefont
  {Yerokhin},\ and\ \citenamefont {Zolotorev}}]{BuC20}%
  \BibitemOpen
  \bibfield  {author} {\bibinfo {author} {\bibfnamefont {D.}~\bibnamefont
  {Budker}}, \bibinfo {author} {\bibfnamefont {J.~R.}\ \bibnamefont {Crespo
  López-Urrutia}}, \bibinfo {author} {\bibfnamefont {A.}~\bibnamefont
  {Derevianko}}, \bibinfo {author} {\bibfnamefont {V.~V.}\ \bibnamefont
  {Flambaum}}, \bibinfo {author} {\bibfnamefont {M.~W.}\ \bibnamefont
  {Krasny}}, \bibinfo {author} {\bibfnamefont {A.}~\bibnamefont {Petrenko}},
  \bibinfo {author} {\bibfnamefont {S.}~\bibnamefont {Pustelny}}, \bibinfo
  {author} {\bibfnamefont {A.}~\bibnamefont {Surzhykov}}, \bibinfo {author}
  {\bibfnamefont {V.~A.}\ \bibnamefont {Yerokhin}},\ and\ \bibinfo {author}
  {\bibfnamefont {M.}~\bibnamefont {Zolotorev}},\ }\bibfield  {title} {\bibinfo
  {title} {Atomic physics studies at the {G}amma {F}actory at {CERN}},\ }\href
  {https://doi.org/https://doi.org/10.1002/andp.202000204} {\bibfield
  {journal} {\bibinfo  {journal} {Annalen der Physik}\ }\textbf {\bibinfo
  {volume} {532}},\ \bibinfo {pages} {2000204} (\bibinfo {year}
  {2020})}\BibitemShut {NoStop}%
\bibitem [{\citenamefont {Klasnikov}\ \emph {et~al.}(2002)\citenamefont
  {Klasnikov}, \citenamefont {Artemyev}, \citenamefont {Beier}, \citenamefont
  {Eichler}, \citenamefont {Shabaev},\ and\ \citenamefont {Yerokhin}}]{KlA02}%
  \BibitemOpen
  \bibfield  {author} {\bibinfo {author} {\bibfnamefont {A.~E.}\ \bibnamefont
  {Klasnikov}}, \bibinfo {author} {\bibfnamefont {A.~N.}\ \bibnamefont
  {Artemyev}}, \bibinfo {author} {\bibfnamefont {T.}~\bibnamefont {Beier}},
  \bibinfo {author} {\bibfnamefont {J.}~\bibnamefont {Eichler}}, \bibinfo
  {author} {\bibfnamefont {V.~M.}\ \bibnamefont {Shabaev}},\ and\ \bibinfo
  {author} {\bibfnamefont {V.~A.}\ \bibnamefont {Yerokhin}},\ }\bibfield
  {title} {\bibinfo {title} {Spin-flip process in radiative recombination of an
  electron with {H}- and {L}i-like uranium},\ }\href
  {https://doi.org/10.1103/PhysRevA.66.042711} {\bibfield  {journal} {\bibinfo
  {journal} {Phys. Rev. A}\ }\textbf {\bibinfo {volume} {66}},\ \bibinfo
  {pages} {042711} (\bibinfo {year} {2002})}\BibitemShut {NoStop}%
\bibitem [{\citenamefont {Bondarevskaya}\ \emph {et~al.}(2011)\citenamefont
  {Bondarevskaya}, \citenamefont {Prozorov}, \citenamefont {Labzowsky},
  \citenamefont {Plunien}, \citenamefont {Liesen},\ and\ \citenamefont
  {Bosch}}]{BoP11}%
  \BibitemOpen
  \bibfield  {author} {\bibinfo {author} {\bibfnamefont {A.}~\bibnamefont
  {Bondarevskaya}}, \bibinfo {author} {\bibfnamefont {A.}~\bibnamefont
  {Prozorov}}, \bibinfo {author} {\bibfnamefont {L.}~\bibnamefont {Labzowsky}},
  \bibinfo {author} {\bibfnamefont {G.}~\bibnamefont {Plunien}}, \bibinfo
  {author} {\bibfnamefont {D.}~\bibnamefont {Liesen}},\ and\ \bibinfo {author}
  {\bibfnamefont {F.}~\bibnamefont {Bosch}},\ }\bibfield  {title} {\bibinfo
  {title} {Theory of the polarization of highly charged ions in storage rings:
  {P}roduction, preservation, observation and application to the search for a
  violation of the fundamental symmetries},\ }\href
  {https://doi.org/https://doi.org/10.1016/j.physrep.2011.06.001} {\bibfield
  {journal} {\bibinfo  {journal} {Physics Reports}\ }\textbf {\bibinfo {volume}
  {507}},\ \bibinfo {pages} {1 } (\bibinfo {year} {2011})}\BibitemShut
  {NoStop}%
\bibitem [{\citenamefont {Bliokh}\ \emph {et~al.}(2007)\citenamefont {Bliokh},
  \citenamefont {Bliokh}, \citenamefont {Savel'ev},\ and\ \citenamefont
  {Nori}}]{BlB07}%
  \BibitemOpen
  \bibfield  {author} {\bibinfo {author} {\bibfnamefont {K.~Y.}\ \bibnamefont
  {Bliokh}}, \bibinfo {author} {\bibfnamefont {Y.~P.}\ \bibnamefont {Bliokh}},
  \bibinfo {author} {\bibfnamefont {S.}~\bibnamefont {Savel'ev}},\ and\
  \bibinfo {author} {\bibfnamefont {F.}~\bibnamefont {Nori}},\ }\bibfield
  {title} {\bibinfo {title} {Semiclassical dynamics of electron wave packet
  states with phase vortices},\ }\href
  {https://doi.org/10.1103/PhysRevLett.99.190404} {\bibfield  {journal}
  {\bibinfo  {journal} {Phys. Rev. Lett.}\ }\textbf {\bibinfo {volume} {99}},\
  \bibinfo {pages} {190404} (\bibinfo {year} {2007})}\BibitemShut {NoStop}%
\bibitem [{\citenamefont {Uchida}\ and\ \citenamefont
  {Tonomura}(2010)}]{UcT10}%
  \BibitemOpen
  \bibfield  {author} {\bibinfo {author} {\bibfnamefont {M.}~\bibnamefont
  {Uchida}}\ and\ \bibinfo {author} {\bibfnamefont {A.}~\bibnamefont
  {Tonomura}},\ }\bibfield  {title} {\bibinfo {title} {Generation of electron
  beams carrying orbital angular momentum},\ }\href
  {https://doi.org/10.1038/nature08904} {\bibfield  {journal} {\bibinfo
  {journal} {Nature}\ }\textbf {\bibinfo {volume} {464}},\ \bibinfo {pages}
  {737} (\bibinfo {year} {2010})}\BibitemShut {NoStop}%
\bibitem [{\citenamefont {Lloyd}\ \emph {et~al.}(2017)\citenamefont {Lloyd},
  \citenamefont {Babiker}, \citenamefont {Thirunavukkarasu},\ and\
  \citenamefont {Yuan}}]{LlB17}%
  \BibitemOpen
  \bibfield  {author} {\bibinfo {author} {\bibfnamefont {S.~M.}\ \bibnamefont
  {Lloyd}}, \bibinfo {author} {\bibfnamefont {M.}~\bibnamefont {Babiker}},
  \bibinfo {author} {\bibfnamefont {G.}~\bibnamefont {Thirunavukkarasu}},\ and\
  \bibinfo {author} {\bibfnamefont {J.}~\bibnamefont {Yuan}},\ }\bibfield
  {title} {\bibinfo {title} {Electron vortices: {B}eams with orbital angular
  momentum},\ }\href {https://doi.org/10.1103/RevModPhys.89.035004} {\bibfield
  {journal} {\bibinfo  {journal} {Rev. Mod. Phys.}\ }\textbf {\bibinfo {volume}
  {89}},\ \bibinfo {pages} {035004} (\bibinfo {year} {2017})}\BibitemShut
  {NoStop}%
\bibitem [{\citenamefont {Bliokh}\ \emph {et~al.}(2017)\citenamefont {Bliokh},
  \citenamefont {Ivanov}, \citenamefont {Guzzinati}, \citenamefont {Clark},
  \citenamefont {{Van Boxem}}, \citenamefont {B\'ech\'e}, \citenamefont
  {Juchtmans}, \citenamefont {Alonso}, \citenamefont {Schattschneider},
  \citenamefont {Nori},\ and\ \citenamefont {Verbeeck}}]{BlI17}%
  \BibitemOpen
  \bibfield  {author} {\bibinfo {author} {\bibfnamefont {K.}~\bibnamefont
  {Bliokh}}, \bibinfo {author} {\bibfnamefont {I.}~\bibnamefont {Ivanov}},
  \bibinfo {author} {\bibfnamefont {G.}~\bibnamefont {Guzzinati}}, \bibinfo
  {author} {\bibfnamefont {L.}~\bibnamefont {Clark}}, \bibinfo {author}
  {\bibfnamefont {R.}~\bibnamefont {{Van Boxem}}}, \bibinfo {author}
  {\bibfnamefont {A.}~\bibnamefont {B\'ech\'e}}, \bibinfo {author}
  {\bibfnamefont {R.}~\bibnamefont {Juchtmans}}, \bibinfo {author}
  {\bibfnamefont {M.}~\bibnamefont {Alonso}}, \bibinfo {author} {\bibfnamefont
  {P.}~\bibnamefont {Schattschneider}}, \bibinfo {author} {\bibfnamefont
  {F.}~\bibnamefont {Nori}},\ and\ \bibinfo {author} {\bibfnamefont
  {J.}~\bibnamefont {Verbeeck}},\ }\bibfield  {title} {\bibinfo {title} {Theory
  and applications of free-electron vortex states},\ }\href
  {https://doi.org/https://doi.org/10.1016/j.physrep.2017.05.006} {\bibfield
  {journal} {\bibinfo  {journal} {Physics Reports}\ }\textbf {\bibinfo {volume}
  {690}},\ \bibinfo {pages} {1} (\bibinfo {year} {2017})}\BibitemShut {NoStop}%
\bibitem [{\citenamefont {Grillo}\ \emph {et~al.}(2017)\citenamefont {Grillo},
  \citenamefont {Harvey}, \citenamefont {Venturi}, \citenamefont {Pierce},
  \citenamefont {Balboni}, \citenamefont {Bouchard}, \citenamefont
  {Carlo~Gazzadi}, \citenamefont {Frabboni}, \citenamefont {Tavabi},
  \citenamefont {Li}, \citenamefont {Dunin-Borkowski}, \citenamefont {Boyd},
  \citenamefont {McMorran},\ and\ \citenamefont {Karimi}}]{GrH17}%
  \BibitemOpen
  \bibfield  {author} {\bibinfo {author} {\bibfnamefont {V.}~\bibnamefont
  {Grillo}}, \bibinfo {author} {\bibfnamefont {T.~R.}\ \bibnamefont {Harvey}},
  \bibinfo {author} {\bibfnamefont {F.}~\bibnamefont {Venturi}}, \bibinfo
  {author} {\bibfnamefont {J.~S.}\ \bibnamefont {Pierce}}, \bibinfo {author}
  {\bibfnamefont {R.}~\bibnamefont {Balboni}}, \bibinfo {author} {\bibfnamefont
  {F.}~\bibnamefont {Bouchard}}, \bibinfo {author} {\bibfnamefont
  {G.}~\bibnamefont {Carlo~Gazzadi}}, \bibinfo {author} {\bibfnamefont
  {S.}~\bibnamefont {Frabboni}}, \bibinfo {author} {\bibfnamefont {A.~H.}\
  \bibnamefont {Tavabi}}, \bibinfo {author} {\bibfnamefont {Z.-A.}\
  \bibnamefont {Li}}, \bibinfo {author} {\bibfnamefont {R.~E.}\ \bibnamefont
  {Dunin-Borkowski}}, \bibinfo {author} {\bibfnamefont {R.~W.}\ \bibnamefont
  {Boyd}}, \bibinfo {author} {\bibfnamefont {B.~J.}\ \bibnamefont {McMorran}},\
  and\ \bibinfo {author} {\bibfnamefont {E.}~\bibnamefont {Karimi}},\
  }\bibfield  {title} {\bibinfo {title} {Observation of nanoscale magnetic
  fields using twisted electron beams},\ }\href@noop {} {\bibfield  {journal}
  {\bibinfo  {journal} {Nature Communications}\ }\textbf {\bibinfo {volume}
  {8}},\ \bibinfo {pages} {689} (\bibinfo {year} {2017})}\BibitemShut {NoStop}%
\bibitem [{\citenamefont {Ivanov}\ and\ \citenamefont
  {Karlovets}(2013)}]{IvK13}%
  \BibitemOpen
  \bibfield  {author} {\bibinfo {author} {\bibfnamefont {I.~P.}\ \bibnamefont
  {Ivanov}}\ and\ \bibinfo {author} {\bibfnamefont {D.~V.}\ \bibnamefont
  {Karlovets}},\ }\bibfield  {title} {\bibinfo {title} {Detecting transition
  radiation from a magnetic moment},\ }\href
  {https://doi.org/10.1103/PhysRevLett.110.264801} {\bibfield  {journal}
  {\bibinfo  {journal} {Phys. Rev. Lett.}\ }\textbf {\bibinfo {volume} {110}},\
  \bibinfo {pages} {264801} (\bibinfo {year} {2013})}\BibitemShut {NoStop}%
\bibitem [{\citenamefont {Silenko}\ \emph {et~al.}(2018)\citenamefont
  {Silenko}, \citenamefont {Zhang},\ and\ \citenamefont {Zou}}]{SiZ18}%
  \BibitemOpen
  \bibfield  {author} {\bibinfo {author} {\bibfnamefont {A.~J.}\ \bibnamefont
  {Silenko}}, \bibinfo {author} {\bibfnamefont {P.}~\bibnamefont {Zhang}},\
  and\ \bibinfo {author} {\bibfnamefont {L.}~\bibnamefont {Zou}},\ }\bibfield
  {title} {\bibinfo {title} {Relativistic quantum dynamics of twisted electron
  beams in arbitrary electric and magnetic fields},\ }\href
  {https://doi.org/10.1103/PhysRevLett.121.043202} {\bibfield  {journal}
  {\bibinfo  {journal} {Phys. Rev. Lett.}\ }\textbf {\bibinfo {volume} {121}},\
  \bibinfo {pages} {043202} (\bibinfo {year} {2018})}\BibitemShut {NoStop}%
\bibitem [{\citenamefont {Karlovets}\ \emph {et~al.}(2015)\citenamefont
  {Karlovets}, \citenamefont {Kotkin},\ and\ \citenamefont {Serbo}}]{KaK15}%
  \BibitemOpen
  \bibfield  {author} {\bibinfo {author} {\bibfnamefont {D.~V.}\ \bibnamefont
  {Karlovets}}, \bibinfo {author} {\bibfnamefont {G.~L.}\ \bibnamefont
  {Kotkin}},\ and\ \bibinfo {author} {\bibfnamefont {V.~G.}\ \bibnamefont
  {Serbo}},\ }\bibfield  {title} {\bibinfo {title} {Scattering of wave packets
  on atoms in the {B}orn approximation},\ }\href
  {https://doi.org/10.1103/PhysRevA.92.052703} {\bibfield  {journal} {\bibinfo
  {journal} {Phys. Rev. A}\ }\textbf {\bibinfo {volume} {92}},\ \bibinfo
  {pages} {052703} (\bibinfo {year} {2015})}\BibitemShut {NoStop}%
\bibitem [{\citenamefont {Maiorova}\ \emph {et~al.}(2018)\citenamefont
  {Maiorova}, \citenamefont {Fritzsche}, \citenamefont {M\"uller},\ and\
  \citenamefont {Surzhykov}}]{MaF18}%
  \BibitemOpen
  \bibfield  {author} {\bibinfo {author} {\bibfnamefont {A.~V.}\ \bibnamefont
  {Maiorova}}, \bibinfo {author} {\bibfnamefont {S.}~\bibnamefont {Fritzsche}},
  \bibinfo {author} {\bibfnamefont {R.~A.}\ \bibnamefont {M\"uller}},\ and\
  \bibinfo {author} {\bibfnamefont {A.}~\bibnamefont {Surzhykov}},\ }\bibfield
  {title} {\bibinfo {title} {Elastic scattering of twisted electrons by
  diatomic molecules},\ }\href {https://doi.org/10.1103/PhysRevA.98.042701}
  {\bibfield  {journal} {\bibinfo  {journal} {Phys. Rev. A}\ }\textbf {\bibinfo
  {volume} {98}},\ \bibinfo {pages} {042701} (\bibinfo {year}
  {2018})}\BibitemShut {NoStop}%
\bibitem [{\citenamefont {Groshev}\ \emph {et~al.}(2020)\citenamefont
  {Groshev}, \citenamefont {Zaytsev}, \citenamefont {Yerokhin},\ and\
  \citenamefont {Shabaev}}]{GoZ20}%
  \BibitemOpen
  \bibfield  {author} {\bibinfo {author} {\bibfnamefont {M.~E.}\ \bibnamefont
  {Groshev}}, \bibinfo {author} {\bibfnamefont {V.~A.}\ \bibnamefont
  {Zaytsev}}, \bibinfo {author} {\bibfnamefont {V.~A.}\ \bibnamefont
  {Yerokhin}},\ and\ \bibinfo {author} {\bibfnamefont {V.~M.}\ \bibnamefont
  {Shabaev}},\ }\bibfield  {title} {\bibinfo {title} {Bremsstrahlung from
  twisted electrons in the field of heavy nuclei},\ }\href
  {https://doi.org/10.1103/PhysRevA.101.012708} {\bibfield  {journal} {\bibinfo
   {journal} {Phys. Rev. A}\ }\textbf {\bibinfo {volume} {101}},\ \bibinfo
  {pages} {012708} (\bibinfo {year} {2020})}\BibitemShut {NoStop}%
\bibitem [{\citenamefont {Zaytsev}\ \emph {et~al.}(2017)\citenamefont
  {Zaytsev}, \citenamefont {Serbo},\ and\ \citenamefont {Shabaev}}]{ZaS17}%
  \BibitemOpen
  \bibfield  {author} {\bibinfo {author} {\bibfnamefont {V.~A.}\ \bibnamefont
  {Zaytsev}}, \bibinfo {author} {\bibfnamefont {V.~G.}\ \bibnamefont {Serbo}},\
  and\ \bibinfo {author} {\bibfnamefont {V.~M.}\ \bibnamefont {Shabaev}},\
  }\bibfield  {title} {\bibinfo {title} {Radiative recombination of twisted
  electrons with bare nuclei: {G}oing beyond the {B}orn approximation},\ }\href
  {https://doi.org/10.1103/PhysRevA.95.012702} {\bibfield  {journal} {\bibinfo
  {journal} {Phys. Rev. A}\ }\textbf {\bibinfo {volume} {95}},\ \bibinfo
  {pages} {012702} (\bibinfo {year} {2017})}\BibitemShut {NoStop}%
\bibitem [{\citenamefont {Maiorova}\ \emph {et~al.}(2020)\citenamefont
  {Maiorova}, \citenamefont {Peshkov},\ and\ \citenamefont
  {Surzhykov}}]{MaP20}%
  \BibitemOpen
  \bibfield  {author} {\bibinfo {author} {\bibfnamefont {A.~V.}\ \bibnamefont
  {Maiorova}}, \bibinfo {author} {\bibfnamefont {A.~A.}\ \bibnamefont
  {Peshkov}},\ and\ \bibinfo {author} {\bibfnamefont {A.}~\bibnamefont
  {Surzhykov}},\ }\bibfield  {title} {\bibinfo {title} {Structured ion beams
  produced by radiative recombination of twisted electrons},\ }\href
  {https://doi.org/10.1103/PhysRevA.101.062704} {\bibfield  {journal} {\bibinfo
   {journal} {Phys. Rev. A}\ }\textbf {\bibinfo {volume} {101}},\ \bibinfo
  {pages} {062704} (\bibinfo {year} {2020})}\BibitemShut {NoStop}%
\bibitem [{\citenamefont {Surzhykov}\ \emph {et~al.}(2008)\citenamefont
  {Surzhykov}, \citenamefont {Jentschura}, \citenamefont {Stöhlker},\ and\
  \citenamefont {Fritzsche}}]{SuJ08}%
  \BibitemOpen
  \bibfield  {author} {\bibinfo {author} {\bibfnamefont {A.}~\bibnamefont
  {Surzhykov}}, \bibinfo {author} {\bibfnamefont {U.~D.}\ \bibnamefont
  {Jentschura}}, \bibinfo {author} {\bibfnamefont {T.}~\bibnamefont
  {Stöhlker}},\ and\ \bibinfo {author} {\bibfnamefont {S.}~\bibnamefont
  {Fritzsche}},\ }\bibfield  {title} {\bibinfo {title} {Electron capture into
  few-electron heavy ions: {I}ndependent particle model},\ }\href
  {https://doi.org/10.1140/epjd/e2007-00269-3} {\bibfield  {journal} {\bibinfo
  {journal} {The European Physical Journal D}\ }\textbf {\bibinfo {volume}
  {46}},\ \bibinfo {pages} {27} (\bibinfo {year} {2008})}\BibitemShut {NoStop}%
\bibitem [{\citenamefont {Surzhykov}\ \emph {et~al.}(2002)\citenamefont
  {Surzhykov}, \citenamefont {Fritzsche},\ and\ \citenamefont
  {St\"ohlker}}]{SuF02}%
  \BibitemOpen
  \bibfield  {author} {\bibinfo {author} {\bibfnamefont {A.}~\bibnamefont
  {Surzhykov}}, \bibinfo {author} {\bibfnamefont {S.}~\bibnamefont
  {Fritzsche}},\ and\ \bibinfo {author} {\bibfnamefont {T.}~\bibnamefont
  {St\"ohlker}},\ }\bibfield  {title} {\bibinfo {title} {Photon-photon angular
  correlations in the radiative recombination of bare high-{Z} ions},\ }\href
  {https://doi.org/10.1088/0953-4075/35/17/308} {\bibfield  {journal} {\bibinfo
   {journal} {Journal of Physics B: Atomic, Molecular and Optical Physics}\
  }\textbf {\bibinfo {volume} {35}},\ \bibinfo {pages} {3713} (\bibinfo {year}
  {2002})}\BibitemShut {NoStop}%
\bibitem [{\citenamefont {Fano}\ and\ \citenamefont {Racah}(1959)}]{FaR59}%
  \BibitemOpen
  \bibfield  {author} {\bibinfo {author} {\bibfnamefont {U.}~\bibnamefont
  {Fano}}\ and\ \bibinfo {author} {\bibfnamefont {G.}~\bibnamefont {Racah}},\
  }\href {https://books.google.de/books?id=buBEAAAAIAAJ} {\emph {\bibinfo
  {title} {Irreducible tensorial sets}}}\ (\bibinfo  {publisher} {Academic
  Press},\ \bibinfo {year} {1959})\BibitemShut {NoStop}%
\bibitem [{\citenamefont {Blum}(2012)}]{Blu12}%
  \BibitemOpen
  \bibfield  {author} {\bibinfo {author} {\bibfnamefont {K.}~\bibnamefont
  {Blum}},\ }\href {https://doi.org/10.1007/978-3-642-20561-3} {\emph {\bibinfo
  {title} {{Density matrix theory and applications}}}},\ Springer Series on
  Atomic Optical and Plasma Physics\ (\bibinfo  {publisher} {Springer},\
  \bibinfo {address} {Berlin},\ \bibinfo {year} {2012})\BibitemShut {NoStop}%
\bibitem [{\citenamefont {Balashov}\ \emph {et~al.}(2000)\citenamefont
  {Balashov}, \citenamefont {Grum-Grzhimailo},\ and\ \citenamefont
  {Kabachnik}}]{BaG00}%
  \BibitemOpen
  \bibfield  {author} {\bibinfo {author} {\bibfnamefont {V.~V.}\ \bibnamefont
  {Balashov}}, \bibinfo {author} {\bibfnamefont {A.~N.}\ \bibnamefont
  {Grum-Grzhimailo}},\ and\ \bibinfo {author} {\bibfnamefont {N.~M.}\
  \bibnamefont {Kabachnik}},\ }\href@noop {} {\emph {\bibinfo {title}
  {Polarization and correlation phenomena in atomic collisions}}}\ (\bibinfo
  {publisher} {Kluwer Academic/Plenum},\ \bibinfo {address} {New York},\
  \bibinfo {year} {2000})\BibitemShut {NoStop}%
\bibitem [{\citenamefont {Prozorov}\ \emph {et~al.}(2003)\citenamefont
  {Prozorov}, \citenamefont {Labzowsky}, \citenamefont {Liesen},\ and\
  \citenamefont {Bosch}}]{PrL03}%
  \BibitemOpen
  \bibfield  {author} {\bibinfo {author} {\bibfnamefont {A.}~\bibnamefont
  {Prozorov}}, \bibinfo {author} {\bibfnamefont {L.}~\bibnamefont {Labzowsky}},
  \bibinfo {author} {\bibfnamefont {D.}~\bibnamefont {Liesen}},\ and\ \bibinfo
  {author} {\bibfnamefont {F.}~\bibnamefont {Bosch}},\ }\bibfield  {title}
  {\bibinfo {title} {Schemes for radiative polarization of ion beams in storage
  rings},\ }\href
  {https://doi.org/https://doi.org/10.1016/j.physletb.2003.09.025} {\bibfield
  {journal} {\bibinfo  {journal} {Physics Letters B}\ }\textbf {\bibinfo
  {volume} {574}},\ \bibinfo {pages} {180 } (\bibinfo {year}
  {2003})}\BibitemShut {NoStop}%
\bibitem [{\citenamefont {Karlovets}\ \emph {et~al.}(2017)\citenamefont
  {Karlovets}, \citenamefont {Kotkin}, \citenamefont {Serbo},\ and\
  \citenamefont {Surzhykov}}]{KaK17}%
  \BibitemOpen
  \bibfield  {author} {\bibinfo {author} {\bibfnamefont {D.~V.}\ \bibnamefont
  {Karlovets}}, \bibinfo {author} {\bibfnamefont {G.~L.}\ \bibnamefont
  {Kotkin}}, \bibinfo {author} {\bibfnamefont {V.~G.}\ \bibnamefont {Serbo}},\
  and\ \bibinfo {author} {\bibfnamefont {A.}~\bibnamefont {Surzhykov}},\
  }\bibfield  {title} {\bibinfo {title} {Scattering of twisted electron wave
  packets by atoms in the {B}orn approximation},\ }\href
  {https://doi.org/10.1103/PhysRevA.95.032703} {\bibfield  {journal} {\bibinfo
  {journal} {Phys. Rev. A}\ }\textbf {\bibinfo {volume} {95}},\ \bibinfo
  {pages} {032703} (\bibinfo {year} {2017})}\BibitemShut {NoStop}%
\bibitem [{\citenamefont {Floettmann}\ and\ \citenamefont
  {Karlovets}(2020)}]{FlK20}%
  \BibitemOpen
  \bibfield  {author} {\bibinfo {author} {\bibfnamefont {K.}~\bibnamefont
  {Floettmann}}\ and\ \bibinfo {author} {\bibfnamefont {D.}~\bibnamefont
  {Karlovets}},\ }\bibfield  {title} {\bibinfo {title} {Quantum mechanical
  formulation of the {B}usch theorem},\ }\href
  {https://doi.org/10.1103/PhysRevA.102.043517} {\bibfield  {journal} {\bibinfo
   {journal} {Phys. Rev. A}\ }\textbf {\bibinfo {volume} {102}},\ \bibinfo
  {pages} {043517} (\bibinfo {year} {2020})}\BibitemShut {NoStop}%
\bibitem [{\citenamefont {Karlovets}(2015)}]{Kar15}%
  \BibitemOpen
  \bibfield  {author} {\bibinfo {author} {\bibfnamefont {D.~V.}\ \bibnamefont
  {Karlovets}},\ }\bibfield  {title} {\bibinfo {title} {Gaussian and {A}iry
  wave packets of massive particles with orbital angular momentum},\ }\href
  {https://doi.org/10.1103/PhysRevA.91.013847} {\bibfield  {journal} {\bibinfo
  {journal} {Phys. Rev. A}\ }\textbf {\bibinfo {volume} {91}},\ \bibinfo
  {pages} {013847} (\bibinfo {year} {2015})}\BibitemShut {NoStop}%
\end{thebibliography}%

\end{document}